\documentclass[fleqn,usenatbib]{mnras}

\usepackage[T1]{fontenc}
\usepackage{ae,aecompl}

\usepackage{graphicx}	% Including figure files
\usepackage{amsmath}	% Advanced maths commands
\usepackage{amssymb}	% Extra maths symbols

\newcommand{\erf}{\operatorname{erf}}
\newcommand{\spline}{\operatorname{spline}}
\newcommand{\rms}{\operatorname{rms}}

\title[MXXL Lightcone Catalogue]{A Lightcone Catalogue from the Millennium-XXL Simulation}

\author[A. M. J. Smith et al.]
{Alex Smith,$^{1}$\thanks{E-mail: a.m.j.smith@durham.ac.uk}
Shaun Cole,$^{1}$
Carlton Baugh,$^{1}$
Zheng Zheng,$^{2}$
Ra\'{u}l Angulo,$^{3}$
\newauthor
Peder Norberg,$^{1,4}$
and Idit Zehavi$^{5}$
\\
$^{1}$Institute for Computational Cosmology, Dept. of Physics, Univ. of Durham, South Road, Durham DH1 3LE, UK \\
$^{2}$Department of Physics and Astronomy, University of Utah, 115 South 1400 East, Salt Lake City, UT 84112, USA \\
$^{3}$Centro de Estudios de F\'{i}sica del Cosmos de Arag\'{o}n, Plaza San Juan 1, Planta-2, 44001, Teruel, Spain \\
$^{4}$Centre for Extragalactic Astronomy, Dept. of Physics, Univ. of Durham, South Road, Durham DH1 3LE, UK \\
$^{5}$Department of Astronomy, Case Western Reserve University, Cleveland, OH 44106, USA
}

\date{Accepted XXX. Received YYY; in original form ZZZ}

\pubyear{2016}

\begin{document}
\label{firstpage}
\pagerange{\pageref{firstpage}--\pageref{lastpage}}
\maketitle

\begin{abstract}
Future galaxy surveys require realistic mock catalogues to understand and quantify 
systematics in order to make precise cosmological measurements. 
We present a halo lightcone catalogue and halo occupation 
distribution (HOD) galaxy catalogue built using the Millennium-XXL (MXXL) simulation.
The halo catalogue covers the full sky, extending to $z=2.2$ with a mass resolution of 
$\sim 10^{11}h^{-1}{\rm M_\odot}$. We use this to build a galaxy catalogue, which has an 
$r$-band magnitude limit of $r<20.0$,
with a median redshift of $z \sim 0.2$. A Monte Carlo HOD method is used to 
assign galaxies to the halo lightcone catalogue, and
we evolve the HODs to reproduce a target luminosity function;
by construction, the luminosity function of galaxies in the mock is in agreement
with the Sloan Digital Sky Survey (SDSS) at low redshifts and the Galaxy and Mass
Assembly (GAMA) survey at high redshifts. A Monte Carlo method
is used to assign a $^{0.1}(g-r)$ colour to each galaxy, and the colour distribution
of galaxies at different redshifts agrees with measurements from GAMA.
The clustering of galaxies in the mock for galaxies in different magnitude
and redshift bins is in good agreement with measurements from
SDSS and GAMA, and the colour-dependent clustering is in reasonable agreement.
We show that the baryon acoustic oscillation (BAO) can be measured in the
mock catalogue, and the redshift space distortions are in agreement
with measurements from SDSS, illustrating that this catalogue will be 
useful for upcoming surveys. 
%The HOD method outlined could be combined with a 
%fast approximate technique for generating halo catalogues to estimate
%error covariances.
\end{abstract}

\begin{keywords}
methods: analytical -- galaxies: haloes -- dark energy -- dark matter -- large-scale structure of Universe
\end{keywords}

%%%%%%%%%%%%%%%%%%%%%%%%%%%%%%%%%%%%%%%%%%%%%%%%%%

%%%%%%%%%%%%%%%%% BODY OF PAPER %%%%%%%%%%%%%%%%%%

\section{Introduction}

Upcoming galaxy surveys, such as the Dark Energy Spectroscopic
Instrument (DESI) survey \citep{DESI2016a,DESI2016b} and
Euclid \citep{Laureijs2011}, aim to measure
the expansion history of the Universe and the growth of 
cosmic structures. Measurements of galaxy clustering, redshift
space distortions and weak lensing will test general relativity,
constrain theories of dark energy, and give us precise cosmological 
constraints.

In order to reach the high precision required to meet these aims, 
it is necessary to understand and quantify the systematic uncertainties 
in measurements from surveys, which requires the use of accurate mock 
catalogues \citep{Baugh2008}. Since a mock catalogue has
a known cosmology and the `true' value of a statistic can
be measured directly, they can be used to develop and test the
analysis tools which will be used on real observations.

Mocks are also required to test observational strategies and 
quantify the resultant levels of sample incompleteness. It is often not possible
to assign a fibre to every galaxy due to mechanical constraints on fibre positioning
\citep[e.g.][]{Hawkins2003,Guo2012,Burden2016,Hahn2016,Pinol2016}
and even if a fibre is assigned, a redshift measurement can fail if 
the galaxy has weak emission lines or low surface brightness. This 
incompleteness may have a significant effect on clustering
measurements, and therefore in order to make precise baryon acoustic oscillation 
(BAO) and redshift space distortion measurements,
it is important that this incompleteness is well understood, and 
that methods are developed and tested in order to mitigate these effects
on the measured clustering. The differences in the clustering statistics
expected in viable models is small, making it essential that 
systematics like these are understood.

Mock catalogues which have realistic galaxy clustering can be created 
from cosmological simulations. In order to see the
BAO peak in clustering measurements, at a scale of the order of $100 h^{-1} \rm Mpc$, 
these simulations need to have a very large box size of the order of a few Gpc. 
Running a hydrodynamical simulation that has both the large
volume needed to model such scales, and the resolution to produce faint galaxies 
down to the flux limit of the survey is infeasible, due to the large computational 
expense. Dark matter only simulations are much less expensive. 
There are several schemes which can be used to populate haloes in a
dark matter only simulation with galaxies. These include the halo 
occupation distribution (HOD) 
\citep[e.g.][]{Peacock2000,Seljak2000,Scoccimarro2001,Berlind2002,
Kravtsov2004,Zheng2005}, which describes the 
probability a halo with mass $M$ contains $N$ galaxies with some
property; the closely related conditional luminosity function (CLF) 
\citep[e.g.][]{Yang2003}, which specifies the luminosity function of galaxies 
at each halo mass;
subhalo abundance matching (SHAM) \citep[e.g.][]{Vale2004,Conroy2006},
which assumes a correlation between halo or subhalo properties (e.g. mass
or circular velocity), and galaxy properties (e.g. luminosity or
stellar mass); and semi-analytic models (SAMs) 
\citep[e.g.][]{Baugh2006,Benson2010,Somerville2015},
which uses analytic prescriptions to model the formation and evolution
of galaxies.

In order to apply a SAM to a simulation, high resolution merger
trees are needed, and these are difficult to construct for 
large volume simulations. However, there are approaches which
can augment the resolution of the simulation merger trees 
\citep[e.g.][]{delaTorre2013,Angulo2014,Benson2016}.
The SHAM prescription assigns galaxies to subhaloes, requiring
a complete subhalo catalogue. Since subhaloes are disrupted 
when they undergo mergers, this catalogue will only be complete for
large subhaloes with thousands of particles, and so a very high
resolution simulation is needed to resolve the low mass subhaloes that
will be populated by faint galaxies. The HOD, on the other hand, can be 
applied to a lower resolution simulation, since satellite galaxies can be
placed around the central galaxy following an analytic distribution, 
without knowledge of the subhaloes. The HOD method can also be applied
to simulations in which the underlying cosmology has been rescaled 
\citep[e.g.][]{Angulo2010}.

Ideally, these methods would be used to populate a halo lightcone 
that is the direct output from a simulation. However, most simulations
do not output lightcones, but output snapshots at discrete
times. Typically when a HOD method is used, it is applied to
a single snapshot. However, this means that the halo 
bias is constant, and so the clustering of haloes does not evolve
with redshift in the mock. Multiple snapshots can be
joined together to create a lightcone, but this leads to
discontinuities at the boundaries; the same halo could appear twice
at either side of the boundary, or not at all \citep[e.g.][]{Fosalba2015}.

The standard abundance matching and HOD schemes do not incorporate 
evolution. Attempts have been made to extend the abundance matching
scheme, such as \citet{Moster2013}, which reproduces the observed
stellar mass function at different redshifts. There is currently
no complete model for HOD evolution, as this evolution would 
depend on the galaxy sample under consideration. \citet{Contreras2016} 
use the HODs produced in SAMs to build
a simple parametrisation of the evolution of the HOD parameters.

Here, we describe a HOD method which we use to populate
haloes over a range of redshifts from the Millennium-XXL (MXXL) 
simulation with galaxies. We first 
create a halo lightcone catalogue from the simulation by interpolating
the positions of haloes between snapshots, which is then populated
with galaxies using HODs, reproducing the observed clustering
from the Sloan Digital Sky Survey (SDSS) \citep{Abazajian2009}
and the Galaxy and Mass Assembly (GAMA) survey 
\citep{Driver2009,Driver2011,Liske2015}.

This paper is organised as follows: in Section~\ref{sec:halo_lightcone} 
we describe the MXXL simulation and outline the method for generating the 
halo lightcone catalogue. In Section~\ref{sec:hods}, we describe the halo 
occupation distribution model, and our method of evolving the HODs with 
redshift. In Section~\ref{sec:mock_catalogue} we outline the method used 
to populate the halo lightcone with galaxies, and the method used to assign 
each galaxy a $^{0.1}(g-r)$ colour. In Section~\ref{sec:applications}, we 
give examples of potential applications of the mock catalogue.

%%%%% SECTION - HALO LIGHTCONE CATALOGUE %%%%%

\section{Halo Lightcone Catalogue} \label{sec:halo_lightcone}

\subsection{The MXXL Simulation}  \label{sec:mxxl}
The Millennium-XXL (MXXL) simulation \citep{Angulo2012b} is a large dark-matter
only N-body simulation in the same family as the Millennium simulation 
\citep{Springel2005}. The volume of MXXL is 216 times larger than Millennium, 
with a box size of $3 h^{-1} \rm Gpc$, and the particle mass is 
$6.17 \times 10^9 h^{-1} \rm M_{\odot}$, with a force softening of $13.7$~kpc. 
MXXL adopts a $\Lambda$CDM cosmology with the same WMAP-1 cosmological parameters 
as the Millennium simulation, $\Omega_{\rm m} = 0.25$, $\Omega_{\Lambda}=0.75$, 
$\sigma_8=0.9$, $h=0.73$, and $n=1$ \citep{Spergel2003}. The initial conditions
were set at a starting redshift of $z=63$, and the simulation was evolved to
$z=0$ with 63 outputs. The large volume of the 
simulation means that it can be used to study features such as baryon acoustic 
oscillations (BAOs) and redshift space distortions with good statistics.

\subsection{Merger Trees}

We use the halo merger trees computed by~\citet{Angulo2012b}. Haloes were found 
using a friends-of-friends (FOF) algorithm~\citep{Davis1985}, and bound subhaloes 
were identified using SUBFIND \citep{Springel2001}. Halo merger trees were built by
identifying the unique descendant of each subhalo at the subsequent snapshot. For
each subhalo, the 15 most bound particles were found, and the subhalo at the next 
snapshot which contains the greatest number of these particles was defined as the 
descendant. In the case that two subhaloes contain equal numbers of these particles, 
the subhalo with the greatest total binding energy was chosen~\citep{Angulo2012a}.

\subsection{Constructing the Halo Lightcone Catalogue}

The full sky halo lightcone catalogue is created using the standard interpolation
method \citep[e.g.][but applied to haloes rather than galaxies]{Merson2013}.
An observer is firstly placed randomly inside the MXXL box. If the observer 
happened to be placed at the centre, haloes at the edge of the box would 
have a redshift $z \sim 0.5$; multiple periodic replications of the box must 
therefore be used in order to construct a catalogue that goes to redshifts higher
than this. This replication is done without any artificial rotation or translation 
in order to prevent the introduction of discontinuities. The positions and velocities of 
each halo at each snapshot are used to interpolate their trajectories through 
the simulation. From the position and redshift of a halo at two adjacent snapshots,
it can be determined whether the halo crossed the observer's lightcone; if it has, a 
binary search algorithm is used to find the interpolated position (and velocity) at
the redshift where it crosses.
    
The halo occupation distribution method of creating the galaxy catalogue
(Section~\ref{sec:hods}) assigns galaxies to FOF groups.
Since the merger tree is defined for SUBFIND subhaloes, we need to infer
the merger tree for the FOF haloes. To do this, we make the assumption that the
position (and velocity) of the main subhalo (i.e. the most massive subhalo) in each 
FOF group is the same
as that of the FOF group itself. The descendant FOF group is then found from the
descendant of the main subhalo. To interpolate the position (and velocity) of
each subhalo, we use cubic interpolation (i.e. use a cubic polynomial to 
describe the path of the halo in each dimension, using the positions and 
velocities at the previous and next snapshot as boundary conditions).

We use a halo mass definition of $M_{200 \rm m}$ (the mass enclosed
by a sphere, centred on the halo, in which the average density is 200 times the 
mean density of the Universe), as stored in the MXXL output for each FOF group. 
Since the number of galaxies in each halo depends on its mass, $M_{200 \rm m}$ 
must be interpolated between snapshots. Below $z=2$, the simulation snapshots
are approximately spaced linearly with expansion factor. We use the 
descendant of the
most massive subhalo to find the descendant of each halo, and then interpolate
linearly in mass between snapshots, finding the mass at the redshift at which
it crosses the lightcone. In the case that two or more haloes merge between 
snapshots, the total mass of the haloes is interpolated linearly, and each halo is 
assigned a constant fraction of the total mass. If the halo is not the most
massive progenitor, a random time between snapshots is chosen
for the merger to take place. If the halo crosses the observer's lightcone
after this time, the merger has happened, and the interpolated mass of the
halo is transferred to the most massive progenitor.
    
\begin{figure}   % HALO MASS FUNCTION
\includegraphics[width=\columnwidth]{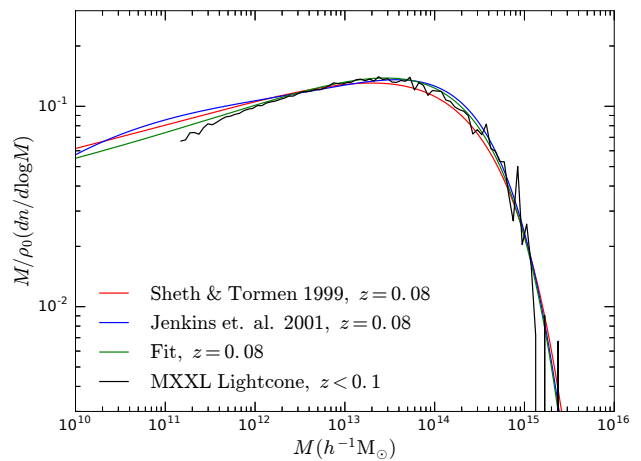}
\caption{Mass function of the halo lightcone catalogue for $z<0.1$, (black), 
compared to the analytic mass function of \citet{Sheth1999} (red), \citet{Jenkins2001} 
(blue), and our fit to the MXXL mass function (green), at the median redshift
$z=0.08$. Halo masses are defined as $M_{200\rm m}$, and have been interpolated 
linearly between simulation snapshots.}
\label{fig:mass_function}
\end{figure}

The mass function of the halo lightcone at low redshifts $z<0.1$ is shown in 
Fig.~\ref{fig:mass_function} and compared to the \citet{Sheth1999} and 
\citet{Jenkins2001} analytic mass functions. At high masses, the mass function of 
the lightcone catalogue is in reasonable agreement with \citet{Sheth1999} and
\citet{Jenkins2001}, but there is a lower abundance of less massive haloes. 
This difference is because haloes in the simulation are identified using a 
FOF algorithm, and not a spherical overdensity (SO) finder. However, a SO mass is
calculated for each FOF halo ($M_{\rm 200m}$). Any small overdensities close
to a large FOF group would be identified as part of the large FOF group, and 
therefore these would be missing from the halo catalogue.

In order to add haloes to the catalogue that are below the MXXL mass
resolution (Section~\ref{sec:add_unresolved_haloes}), 
and to evolve the HODs with redshift (Section~\ref{sec:HOD_evolution}), it 
is useful to have a smooth function
which is in close agreement with the actual mass function of the catalogue.
For this, we take a mass function with the same form as \citet{Sheth1999},
but refit the parameters to the MXXL mass function. This fit is shown
as the green curve in Fig.~\ref{fig:mass_function}, which is in better 
agreement with the MXXL mass function at low masses, and is close to the
fit given in equation~2 of \citet{Angulo2012b}. The MXXL
mass function peels away from this fit slightly at masses close to the resolution
limit, but is complete for masses greater than $\sim 10^{12} h^{-1} \rm M_{\odot}$.
    
\begin{figure}   % HALO NUMBER DENSITY (WITH MASS INTERPOLATION)
\includegraphics[width=\columnwidth]{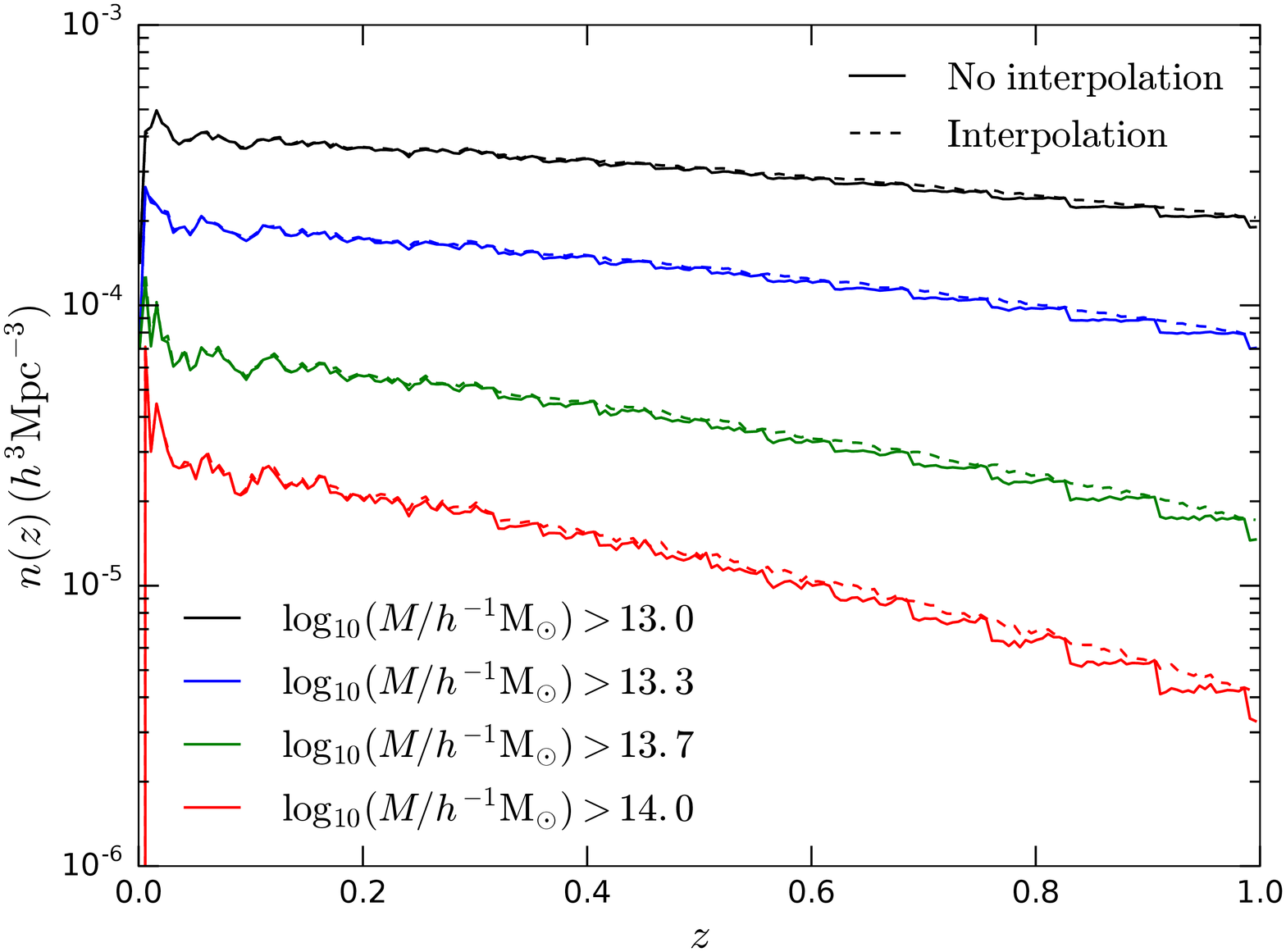}
\caption{Number density of haloes in the halo lightcone catalogue as a function 
of redshift for haloes with mass $M_{200 \rm m}$ greater than several thresholds, as
indicated by the colour. 
Solid lines are where the halo mass has been kept frozen between snapshots, and 
dashed lines are where the mass has been interpolated.}
\label{fig:n_z_interp}
\end{figure}

The number density of haloes as a function of redshift in the halo lightcone 
catalogue is shown in Fig.~\ref{fig:n_z_interp} for several mass thresholds. 
If halo masses are kept fixed between snapshots, step 
features can be seen, since the mass function is being kept frozen and will only 
change at the next snapshot. These features are most apparent for the highest mass
threshold, for which the number density decreases more rapidly at high redshifts. Mass
interpolation greatly reduces these features.
    
\begin{figure}   % HALO CLUSTERING - BAO
\includegraphics[width=\columnwidth]{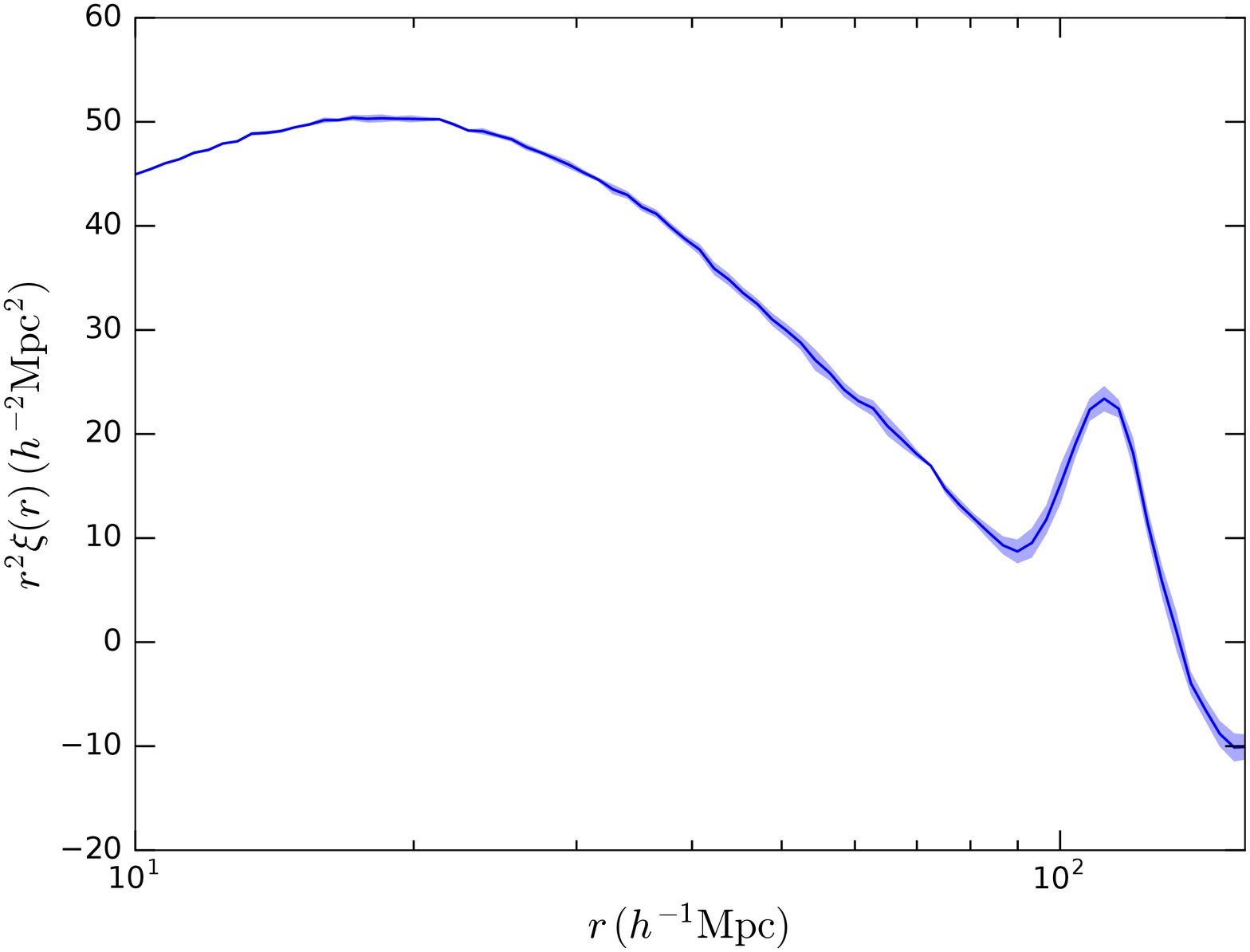}
\caption{Real space correlation function, scaled by $r^2$, of the halo lightcone catalogue 
for haloes with $M_{200 \rm m} > 3 \times 10^{12} h^{-1} \rm M_{\odot}$ and $z<0.5$.
The blue shaded area shows the error on the mean in the clustering, calculated
from four quadrants of the sky.}
\label{fig:xi_bao}
\end{figure}

The large scale real space correlation function of the lightcone catalogue 
for FOF groups with masses $M_{200 \rm m} > 3 \times 10^{12} h^{-1} \rm M_{\odot}$ 
and $z<0.5$ is shown in Fig.~\ref{fig:xi_bao}. This redshift limit avoids structures
being repeated due to periodic replication of the box\footnote{Lightcones with a
wide opening angle, or directed along the principle axes of the simulation, that extend 
beyond $z=0.5$ will contain repeated structures. This will result in clustering errors
being underestimated.}. 
The BAO peak can be seen clearly in the clustering of haloes.

\subsection{Caveats}
	
The interpolation scheme uses the position and velocity of a halo at two snapshots 
as boundary conditions in order to find the path the halo moved through in the
simulation. If two haloes merge together, there is not enough information to determine
when this occurs, so we assume they merge at a random time. If a new halo forms, or drops 
below the resolution limit, we assume this happens exactly on a snapshot.

To construct the merger trees, the descendant of a subhalo is defined as the 
subhalo which contains the majority of its 15 most bound particles
\citep{Angulo2012a}. However, it is likely that some of the particles of the 
descendant subhalo were not in its progenitor, and vice versa. All of these particles 
are used to calculate the position and velocity of the subhalo, which can occasionally
lead to jumps in the position of a subhalo that are inconsistent with its velocity. 

Sometimes, a halo can be lost by the halo finder at one snapshot, but is then found 
again at a later snapshot. This can happen if a small halo passes very close to a 
more massive halo at one snapshot; the SUBFIND algorithm can fail to identify the
small halo as the algorithm finds that its particles are bound to the massive halo. 
The MXXL merger trees we use do not make any attempt to add in these haloes lost by 
SUBFIND. However, since we use the most massive subhalo in a FOF group to trace the 
FOF merger trees, and use $M_{\rm 200m}$ as the mass definition, our results should 
not be affected much by small subhaloes being lost by the halo finder.

\subsection{Haloes below the mass resolution} \label{sec:add_unresolved_haloes}

Populating the resolved haloes in the MXXL halo lightcone with galaxies will
result in incompleteness in a magnitude limited galaxy catalogue at low 
redshifts. This is because intrinsically faint galaxies which are sufficiently 
close to the observer to be bright enough to be included in the catalogue 
occupy haloes which fall below the MXXL mass resolution. We use our fit 
to the MXXL halo mass function in order to add 
these haloes into the lightcone catalogue, and position them randomly
in the catalogue so that they are unclustered. Other methods for augmenting
the halo catalogue exist \citep[e.g.][]{delaTorre2013,Angulo2014,Benson2016}, 
but we find that this simple method is
able to bring the $dN/dz$ of galaxies in the catalogue into better agreement
with the measured $dN/dz$ from GAMA, while only having a very small effect on 
the measured clustering.

The redshift distribution of the haloes which need to be added to the lightcone
catalogue can be calculated from the integral,
\begin{equation} \label{eq:dNdz_unresolved}  % Eq - dN/dz unresolved haloes
\frac{dN}{dz} = \int^{M_{\rm max}}_{M_{\rm min}(z)} n_{\rm unres}(M,z) \frac{dV}{dz} dM,
\end{equation}
where $n_{\rm unres}(M,z)$ is the number density of unresolved haloes, 
$dV/dz$ is the comoving volume per unit redshift, 
$M_{\rm min}(z)$ is the minimum halo mass that could host a galaxy brighter 
than the faintest observable galaxy in the survey at that redshift,
and $M_{\rm max}=10^{12}h^{-1}M_{\odot}$ is the mass at which the
MXXL mass function is judged to be complete.
If the survey is flux limited, then the faintest observable galaxy 
at each redshift is set 
by an apparent magnitude threshold; for our mock catalogue we set
this threshold to $r=20$, as this is the magnitude threshold
for the DESI Bright Galaxy Survey (BGS) \citep{DESI2016a}. 
The number density of unresolved haloes is given by
$n_{\rm unres}(M,z) = n_{\rm fit}(M,z) - n_{\rm res}(M,z)$, where 
$n_{\rm fit}(M,z)$ is our fit to the number density of haloes in the lightcone,
extrapolated to low masses, and $n_{\rm res}(M,z)$ is the number density of haloes 
resolved in MXXL. We model the mass function of resolved haloes by multiplying the 
fit to the mass function by a cutoff at the mass resolution limit of 
$M_{200\rm m} \sim 10^{11} h^{-1}{\rm M_{\odot}}$,
\begin{equation}     % Eq - number density unresolved haloes
n_{\rm unres}(M,z) = [1 - {\rm cut}(M,z)] n_{\rm fit}(M,z),
\end{equation}
where a good approximation to the cutoff is given by
\begin{equation}      % Eq - cutoff in mass function
{\rm cut}(M,z) = 10^{(-z-2)(\log_{10}(M/h^{-1}{\rm M_\odot})-11)^{0.6}}.
\end{equation}

In order to add unresolved haloes to the catalogue, we first randomly
draw a redshift for each unresolved halo from the $dN/dz$ distribution
defined in Eq.~\ref{eq:dNdz_unresolved}. The mass of each halo is 
then randomly assigned using the mass distribution at the redshift of
the halo defined by $n_{\rm unres}(M, z)$. The haloes are then randomly
positioned uniformly on the sky. Since the unresolved haloes are
randomly positioned so that they are unclustered, redshift space
distortions do not affect their clustering, and so we set the velocity of
each of these haloes to zero. A random concentration is also assigned 
from the mass-concentration relation of MXXL (with scatter), extrapolated
to lower masses.

While the introduction of unclustered haloes only has a small effect
on the two-point correlation function, other statistics might also change, for example, 
density estimators and void statistics. We have not checked the size of this
effect, but the final galaxy catalogue includes a
flag which indicates whether a galaxy lives inside one of these haloes,
enabling these galaxies to be removed when calculating other statistics.

%%%%% SECTION - HALO OCCUPATION DISTRIBUTION %%%%%

\section{Halo Occupation Distribution} \label{sec:hods}

Galaxies are biased tracers of the underlying dark matter density field. The halo 
occupation distribution (HOD) describes this bias using the probability that a halo 
of mass $M$ contains $N$ galaxies with a certain property, $P(N|M)$, providing
a physical interpretation of galaxy clustering measurements.
  
The mean number of galaxies in a halo of mass $M$ which are brighter than some 
luminosity threshold, $L$, can be written as a sum of central and satellite galaxies
\citep[e.g.][]{Zheng2005},
\begin{equation} \label{eq:hod_tot}    % Eq - Ngal = sat + cent
\langle N_{\rm gal}(>L|M) \rangle = \langle N_{\rm cen}(>L|M) \rangle + \langle N_{\rm sat} (>L|M) \rangle.
\end{equation}
We use central and satellite occupation functions of the same form as \citet{Zehavi2011}. 
The mean number of central galaxies brighter than $L$ is described by a smoothed step function,
\begin{equation} \label{eq:hod_cent}   % Eq - HOD (cent)
\langle N_{\rm cen}(>L|M) \rangle = \frac{1}{2} \left[ 1 + \erf \left( \frac{\log M - \log M_{\rm min}(L)}{\sigma_{\log M}(L)} \right) \right],
\end{equation}
where $\erf(x) = 2\pi^{-1/2} \int^x_0 e^{-x'^2}dx'$ is the error function. 
The parameter $M_{\rm min}$ is the halo mass for which half of haloes contain a 
galaxy brighter than $L$, and $\sigma_{\log M}$ sets the width of the step. For 
$M \gg M_{\rm min}$, $\langle N_{\rm cen}(>L|M) \rangle = 1$, while for 
$M \ll M_{\rm min}$, $\langle N_{\rm cen}(>L|M) \rangle = 0$. The mean number of 
satellites per halo brighter than $L$ is given by a power law,
\begin{equation} \label{eq:hod_sat}   % Eq - HOD (sat)
\langle N_{\rm sat}(>L|M) \rangle = \langle N_{\rm cen}(>L|M) \rangle \left( \frac{M - M_0(L)}{M'_1(L)} \right)^{\alpha(L)},
\end{equation}
where $M_0$ is the cutoff mass scale, $M'_1$ the normalisation, and $\alpha$ the 
power law slope. $M'_1$ is different to $M_1$, the mass of a halo which on average 
contains 1 satellite, although the two quantities are related\footnote{Since the
satellite occupation function is modified by the centrals, the relation
$M_1 = M'_1 + M_0$ is not exact.}.
The power law is also multiplied by the central occupation function, which
ensures that the brightest galaxy in the halo is the central; there cannot
be a satellite brighter than $L$ without there first being a central galaxy
brighter than $L$. 

\subsection{HODs at redshift 0.1} \label{sec:hods_z0.1}

We use HOD parameters calculated from the SDSS using the procedure of~\citet{Zehavi2011}. 
These are calculated for different luminosity threshold galaxy samples, using an
MCMC code to find the best fitting HOD parameters which reproduce the measured projected 
correlation functions to within the SDSS uncertainties. Since the cosmology of the MXXL 
simulation is different to 
that used by~\citet{Zehavi2011}, the parameter fitting was redone using the 
Millennium cosmology. The SDSS HODs use the absolute
{\it r}-band magnitude of each galaxy, {\it k}-corrected to a reference redshift of 
$z_{\rm ref}=0.1$
(see Section~\ref{sec:k-corrections}), which is the median redshift of the survey.
We denote absolute magnitudes {\it k}-corrected to this redshift as $^{0.1} \! M_r$. 
Absolute magnitudes written as $^{0.1} \! M_r$ assume $h=1$.

\begin{figure}   % HOD PARAMETERS & FITS
\includegraphics[width=\columnwidth]{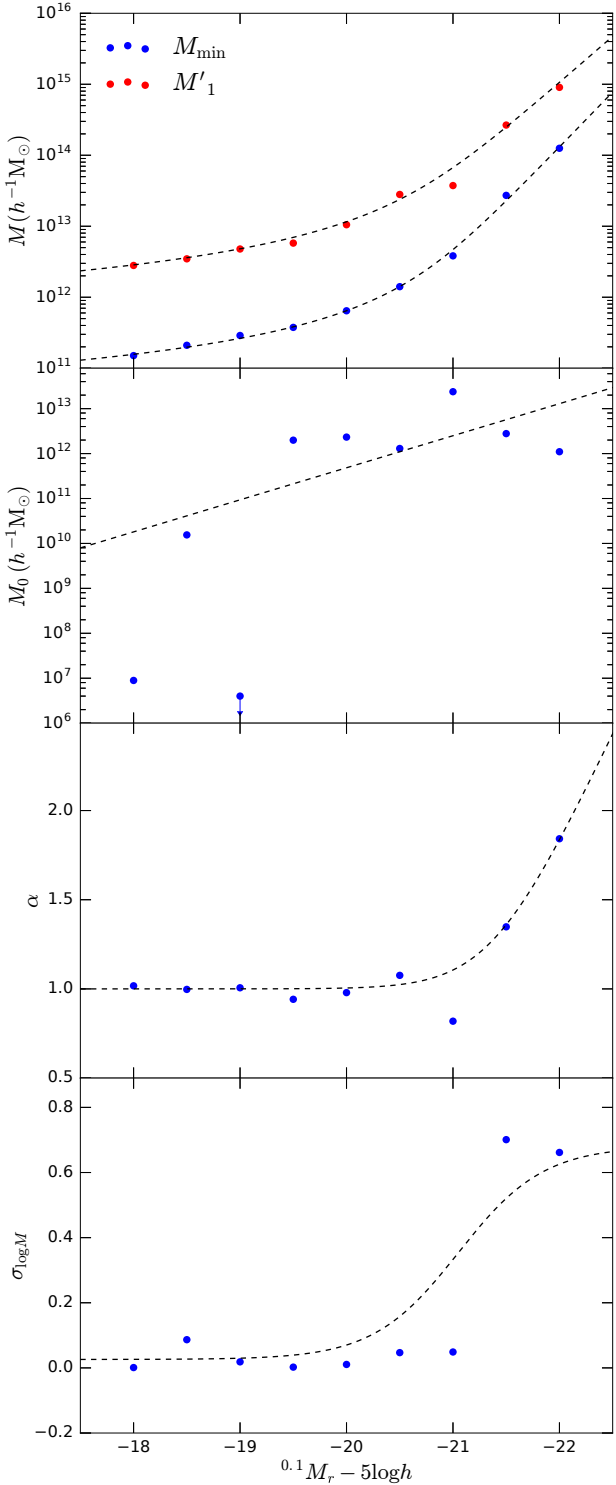}
\caption{Best fitting HOD parameters to the SDSS volume limited samples in
Millennium cosmology (points), and smooth functions fitted to these
points (dashed lines), as a function of magnitude. Top panel: $M_{\rm min}$ 
(blue) and $M'_1$ (red). Second panel: $M_0$. Third panel: $\alpha$. 
Bottom panel: $\sigma_{\log M}$. The $^{0.1}M_r-5\log h=-19$ sample has 
$M_0=10^{-10.2}h^{-1} \rm M_{\odot}$, but $M_0$ is poorly constrained. 
Errors are not shown as they are misleading, due to the highly asymmetric 
probability distributions.}
\label{fig:hod_params}
\end{figure}

The best fitting HOD parameters, in Millennium cosmology, are shown by the points
in Fig~\ref{fig:hod_params}. We do not show the errors as they are misleading, due to 
the probability distributions being highly asymmetric. Projecting these asymmetric probability 
distributions to 1 dimensional errors can lead to the best fitting values of some of the 
parameters being outside the error bars, particularly for $\sigma_{\log M}$.

For each HOD parameter, a least squares routine is used to fit a function which 
describes the variation with luminosity, which are shown by the dashed lines in 
Fig.~\ref{fig:hod_params}. The top panel shows $M_{\rm min}(L)$ and 
$M'_1(L)$, for which we fit curves of the same functional form as Eq.~11 
from \citet{Zehavi2011},
\begin{equation}         % Eq - Fit to Mmin and M1'
L/L_* = A \left( \frac{M}{M_t} \right)^{\alpha_M} \exp \left( -\frac{M_t}{M} +1 \right),
\end{equation}
where $A$, $M_t$ and $\alpha_M$ are free parameters. We fit a power law to 
$M_0(L)$ (second panel). This is a poor fit for the points at $^{0.1}M_r=-18$,
which is over 3 orders of magnitude lower than the fit, and $^{0.1}M_r=-19$, 
which is 20 orders of magnitude lower. However, increasing the value 
of this parameter by many orders of magnitude has a very small effect on the shape 
of the HODs. This is because the occupation function of central galaxies 
adds a second cutoff to Eq.~\ref{eq:hod_sat}; if $M_0$ is below this cutoff,
it will not affect the shape of the HODs. The parameter $\alpha(L)\sim 1$ at low 
luminosities, 
but increases for the highest luminosity samples (third panel). We fit a linear
relation, which smoothly transitions to $\alpha=1$ at low luminosities. 
$\sigma_{\log M}(L)$ (bottom panel) is small at low luminosities, with a step up to 
$\sim 0.7$ for the brightest two samples. We fit a sigmoid function to 
$\sigma_{\log M}$, where the width of the step is set such that the HODs do not overlap.
    
\begin{figure}   % OCCUPATION FUNCTIONS WITH SPLINE KERNEL
\includegraphics[width=\columnwidth]{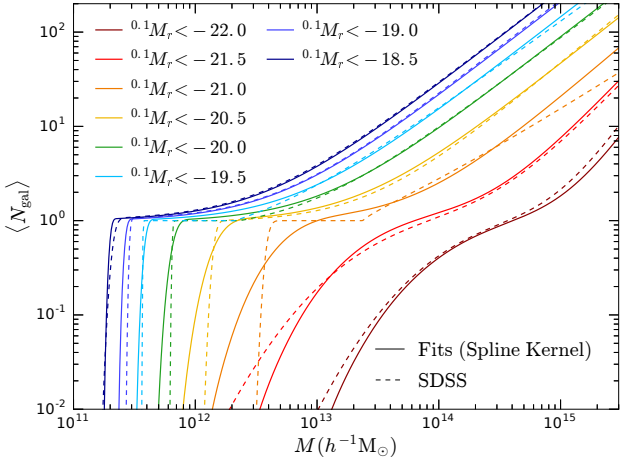}
\caption{Mean halo occupation functions for luminosity threshold samples, as 
described by Eqs.~\ref{eq:hod_tot}-\ref{eq:hod_sat}, using SDSS HOD parameters 
in the Millennium cosmology (dashed lines) and our fits to the HOD parameters, using 
Eq.~\ref{eq:hod_cent_sk} in place of Eq.~\ref{eq:hod_cent} to describe the
contribution from central galaxies (solid lines). Colours indicate the 
luminosity threshold, as shown by the legend.}
\label{fig:HODs}
\end{figure}

The large step in $\sigma_{\log M}(L)$ means that, as the luminosity threshold 
is increased, there is a rapid jump in the amount of scatter in the luminosities 
of central galaxies. This results in overlapping HODs, as can be seen for the 
$^{0.1}M_r < -21$ and $^{0.1}M_r < -21.5$ samples in Fig.~\ref{fig:HODs}. For 
two luminosity thresholds $L_1$ and $L_2$, where $L_1 < L_2$, it must be true that 
$\langle N_{\rm gal} (>L_1|M) \rangle \geq \langle N_{\rm gal} (>L_2|M) \rangle$ 
since all galaxies brighter than $L_2$ are also brighter than $L_1$. However, if 
the two occupation functions cross, then this condition is not satisfied for 
haloes below the mass at which they cross. We therefore must model the HODs 
such that there is no overlap. There exist HOD frameworks in which the occupation
functions cannot overlap~\citep[see e.g.][]{Leauthaud2011}, but since we are using
HOD parameters obtained using the standard HOD framework, we make a small
modification to these HODs to prevent any crossing, as set out below.

Eq.~\ref{eq:hod_cent} assumes that the scatter set by the parameter 
$\sigma_{\log M}(L)$ is Gaussian. Since a Gaussian function has a long tail which 
extends to infinity, there will always be an overlap between HODs if 
$\sigma_{\log M}(L_2) > \sigma_{\log M}(L_1)$. We instead approximate the Gaussian 
by using a spline kernel \citep{Schoenberg1946},
\begin{equation} \label{eq:sk}   % Eq - Spline kernel definition
\spline(x) = 
\begin{array}{lr}
1 - 6|x|^2 + 6|x|^3 & |x| \leq 0.5\\
2(1-|x|)^3          & 0.5 < |x| \leq 1 \\
0                   & |x| > 1,
\end{array}
\end{equation}
which has $\spline(0)=1$, mean = 0, variance = 1/12, and $\spline(x)=0$ for $|x|>1$.
This function can be rescaled and normalised to approximate any Gaussian of mean 
$\mu$ and variance $\sigma^2$ as,
\begin{equation} \label{eq:sk_scaled}  % Eq - Scaled spline kernel
S(x) = \frac{4/3}{\sigma\sqrt{12}} \spline \left( \frac{x-\mu}{\sigma \sqrt{12}}\right).
\end{equation}
The HOD for central galaxies can therefore be written as
\begin{equation} \label{eq:hod_cent_sk}   % Eq - HOD (cent) with spline kernel
\langle N_{\rm cen}(>L|M) \rangle = \frac{1}{2} \left[ 1 + F \left( \frac{\log M - \log M_{\rm min}(L)}{\sigma_{\log M}(L)}\right) \right],
\end{equation}
where $F(x) = 2\int^x_0 S(x')dx'$. The best fitting values of $\sigma_{\log M}$
(shown by the points in the bottom panel of Fig.~\ref{fig:hod_params}), suggest 
a sharp step between $^{0.1}M_r=-21$ and $^{0.1}M_r=-21.5$. Even using
Eq.~\ref{eq:hod_cent_sk}, the HODs will still overlap with this abrupt step,
but they will not overlap if the step is gradual, unlike 
Eq.~\ref{eq:hod_cent}. We make the 
step in $\sigma_{\log M}(L)$ as narrow as we can while preventing the HODs 
crossing (shown by the dashed curve).

The HODs using the SDSS HOD parameters, and our fits, are shown in 
Fig.~\ref{fig:HODs}. Our fits produce halo occupation functions which are in 
reasonable agreement with the SDSS HODs, with the exception of the $^{0.1}M_r < -21$ 
and $^{0.1}M_r < -21.5$ samples, where the width of the step set by the parameter 
$\sigma_{\log M}$ is too broad and narrow respectively. This is necessary to prevent 
the HODs from overlapping. The SDSS HOD for the $^{0.1}M_r<-21$ sample appears to 
have a sharp transition from centrals to satellites, which is due to a large value 
of $M_0$ compared to $M'_1$.

\subsection{Redshift Evolution} \label{sec:HOD_evolution}

In order to evolve the occupation functions with redshift, we first choose a 
target luminosity function, $\phi_{\rm target}(L,z)$, that we would like the 
galaxies in the mock catalogue to reproduce as a function of redshift. This 
luminosity function defines a mapping between a luminosity threshold $L$ at 
redshift $z$, and the number density of galaxies brighter than this, 
$n_{\rm gal}^{\rm target}(>L,z)$.
    
For a given HOD, the number density of galaxies brighter than $L$ can be 
calculated from the integral
\begin{equation}         % Eq - galaxy number density from HOD
n_{\rm gal}(>L,z) = \int n_{\rm halo}(M,z) \langle N(>L|M,z) \rangle dM,
\end{equation}
where $n_{\rm halo}(M,z)$ is the number density of haloes of mass $M$ 
at redshift $z$, and $\langle N(>L|M,z) \rangle$ is the halo occupation 
function at redshift $z$. The HODs must evolve with redshift such that 
the condition $n_{\rm gal}(>L,z) = n_{\rm gal}^{\rm target}(>L, z)$ is 
satisfied. 
    
Since the target luminosity function defines a mapping between a luminosity 
threshold and the number density of galaxies, the occupation functions can 
be rewritten as a function of number density, $n_{\rm gal}$:
$\langle N(>L|M,z) \rangle \equiv \langle N(n_{\rm gal}|M,z) \rangle$. 
The shape of the HOD could evolve in a complex way, but for simplicity we
keep the shape of the occupation function fixed for constant galaxy number 
density, but slide the HODs along the halo mass axis such that the target luminosity function 
is achieved. That is, the HOD parameters $\sigma_{\log M}(n_{\rm gal},z)$ and 
$\alpha(n_{\rm gal},z)$ are kept constant, but the 3 mass parameters 
$M_{\rm min}$, $M_0$ and $M_1$ are all multiplied by some factor $f$,    
\begin{equation}         % Eq - f 'slide' factor
M_{\rm HOD}(n_{\rm gal}, z) = f(n_{\rm gal}, z) M_{\rm HOD}(n_{\rm gal}),
\end{equation}
where $M_{\rm HOD}$ is one of the HOD mass parameters. The value of $f$ required to 
achieve the target luminosity function is found by finding the root
of the equation 
\begin{equation}         % Eq - f root finding
n_{\rm gal}(>L,f(z)) - n_{\rm gal}^{\rm target}(>L,z) = 0.
\end{equation}  
    
%\subsubsection{Target Luminosity Function}
  
At high redshifts, the target luminosity function we use is the evolving
Schechter function fit to the luminosity function estimated from the 
Galaxy and Mass Assembly (GAMA) survey.
The Schechter function can be written in terms of magnitudes as
\begin{equation}       % Eq - Schechter function
\phi(M)=0.4 \ln 10 \, \phi^*(10^{0.4(M^*-M)})^{1+\alpha} \exp (-10^{0.4(M^*-M)}),
\end{equation}
where $\phi^*$ is the normalisation, $M^*$ is a characteristic magnitude 
and $\alpha$ is the faint end slope. For GAMA, \citet{Loveday2012,Loveday2015} model 
the evolution of the Schechter parameters with redshift as
\begin{eqnarray}       % Eq - Schechter parameter evolution
\nonumber \alpha(z) &=& \alpha(z_0)\\
\nonumber M^*(z)    &=& M^*(z_0) - Q(z-z_0)\\
\phi^*(z) &=& \phi^*(0) 10^{0.4Pz},
\end{eqnarray}
where $Q$ parametrises the evolution in luminosity, $P$ parametrises 
the evolution in number density, and $z_0=0.1$ is the same reference 
redshift as used for the {\it k}-corrections (Section~\ref{sec:k-corrections}). 
The faint end slope is kept constant with redshift since there is not 
enough data to constrain it at high redshifts. We use the evolving 
Schechter function, $\phi_{\rm GAMA}$, from~\citet{Loveday2012} with
$P=1.8$ and $Q=0.7$.

However, the shape of the GAMA Schechter luminosity function is slightly 
different than the SDSS luminosity function. Using it as the target
at all redshifts would result in the evolution parameter $f \neq 1$ at $z=0.1$, 
meaning that the HODs would change from the HODs measured from SDSS. In order 
to not change the HODs at $z=0.1$, we use the luminosity function from SDSS,
$\phi_{\rm SDSS}$, as the target at low redshifts. The SDSS target
luminosity function we use is the result of the integral
\begin{equation}
\phi_{\rm SDSS}(>L) = \int n_{\rm halo}(M)\langle N(>L|M) \rangle dM,
\end{equation}
where $n_{\rm halo}(M)$ is the number density of haloes at $z=0.1$,
and $\langle N(>L|M) \rangle$ is the (unevolved) occupation function.
The result of this integral is close to the \citet{Blanton2003}
luminosity function for absolute magnitudes brighter than 
$^{0.1}M_r=-19$, and by definition $f=1$, so 
the HODs remain unchanged from SDSS at this redshift. However, at 
magnitudes fainter
than $^{0.1}M_r=-19$, the result of this integral is very flat, while 
the \citet{Blanton2003} SDSS luminosity function is steeper; at 
$^{0.1}M_r=-17$ they differ by a factor of $\sim 2$.
We therefore smoothly
transition to the \citet{Blanton2003} luminosity function at 
$^{0.1}M_r=-19$. This is then extrapolated to fainter magnitudes
with a power law.

We interpolate the target luminosity function from $\phi_{\rm SDSS}$ at low 
redshifts to $\phi_{\rm GAMA}$ at high redshifts,
\begin{eqnarray}       % Eq - Luminosity function merge
\nonumber
\phi_{\rm target}(M,z) &=& (1-w(z))\phi_{\rm SDSS}(M,z) \\
&+& w(z)\phi_{\rm GAMA}(M,z),
\end{eqnarray}
where the transition between $0.1 < z < 0.2$ is set by the sigmoid
function
\begin{equation}       % Eq - w(z) Luminosity function transition
w(z) = (1 + e^{-100(z-0.15)})^{-1}.
\end{equation}

\begin{figure}   % F VALUES
\includegraphics[width=\columnwidth]{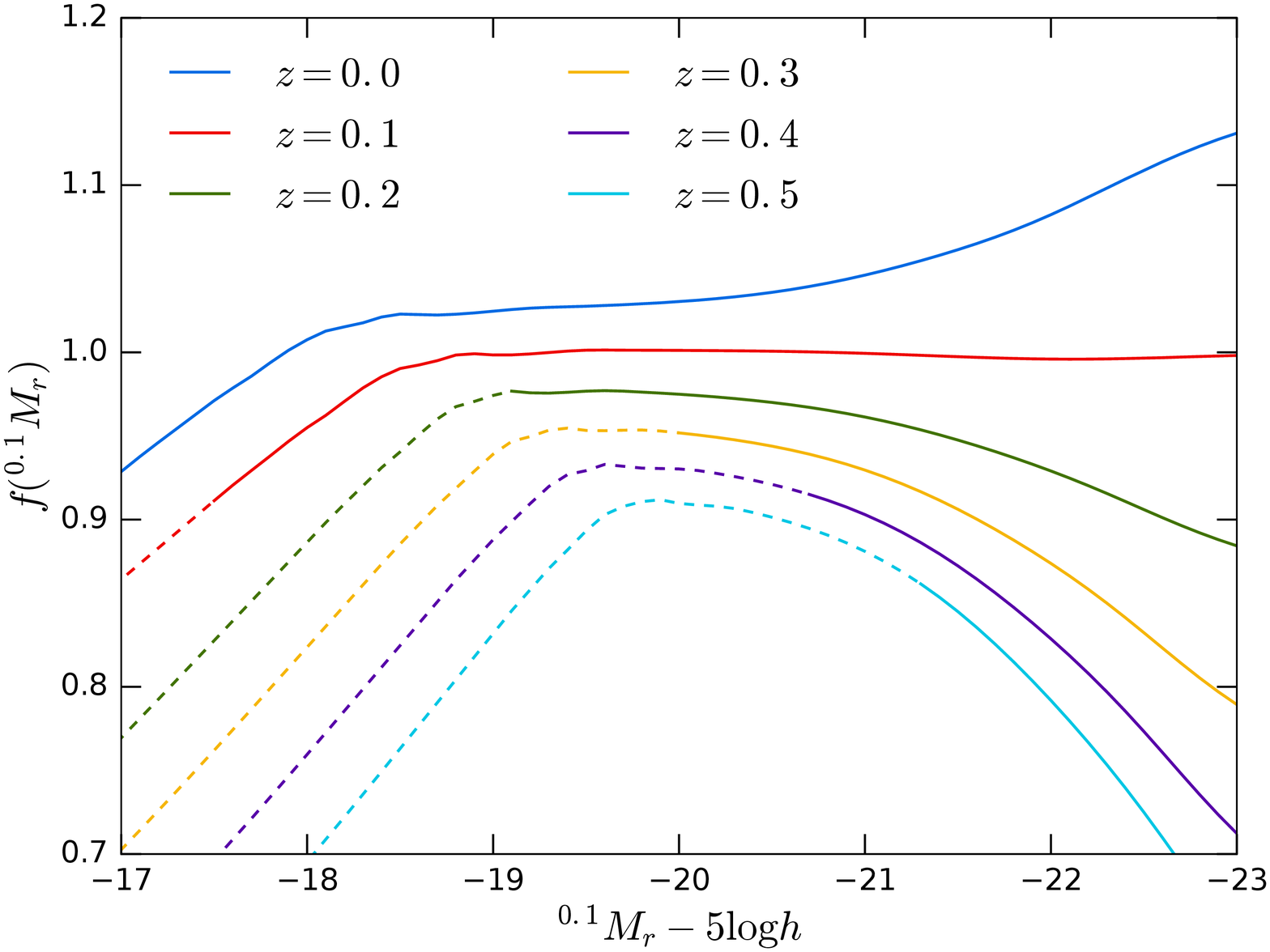}
\caption{Evolution parameter, $f$, as a function of magnitude for different
redshifts, as indicated by the colour. This is the factor by which the HOD 
mass parameters are multiplied
in order to achieve the galaxy number density set by the target 
luminosity function. Dashed lines indicate absolute magnitudes which 
correspond to apparent magnitudes that are fainter 
than the $r=20$ limit at that redshift.}
\label{fig:f_values}
\end{figure}

The evolution parameter, $f$, for this target luminosity function, is shown
in Fig.~\ref{fig:f_values} as a function of magnitude for different redshifts.
At $z=0.1$, $f$ is close to 1, by definition. However, it is not exactly 1
because the function $w(z)$, which sets the transition between the two target 
luminosity functions is close to, but not exactly 0 at $z=0.1$. At $z=0.1$,
$f$ is equal to 1 to within 1\%. Fainter than magnitude $-19$, $f(z=0.1)<1$. 
At these faint magnitudes, the target luminosity function is transitioning
to the \citet{Blanton2003} luminosity function. Keeping $f(z=0.1)=1$ at
all magnitudes produces a luminosity function which, while being close
to SDSS at the bright end, is too flat at the faint end, so this transition
is required to bring the luminosity function of the mock into better agreement
with the data.

\begin{figure}   % M1 EVOLUTION
\includegraphics[width=\columnwidth]{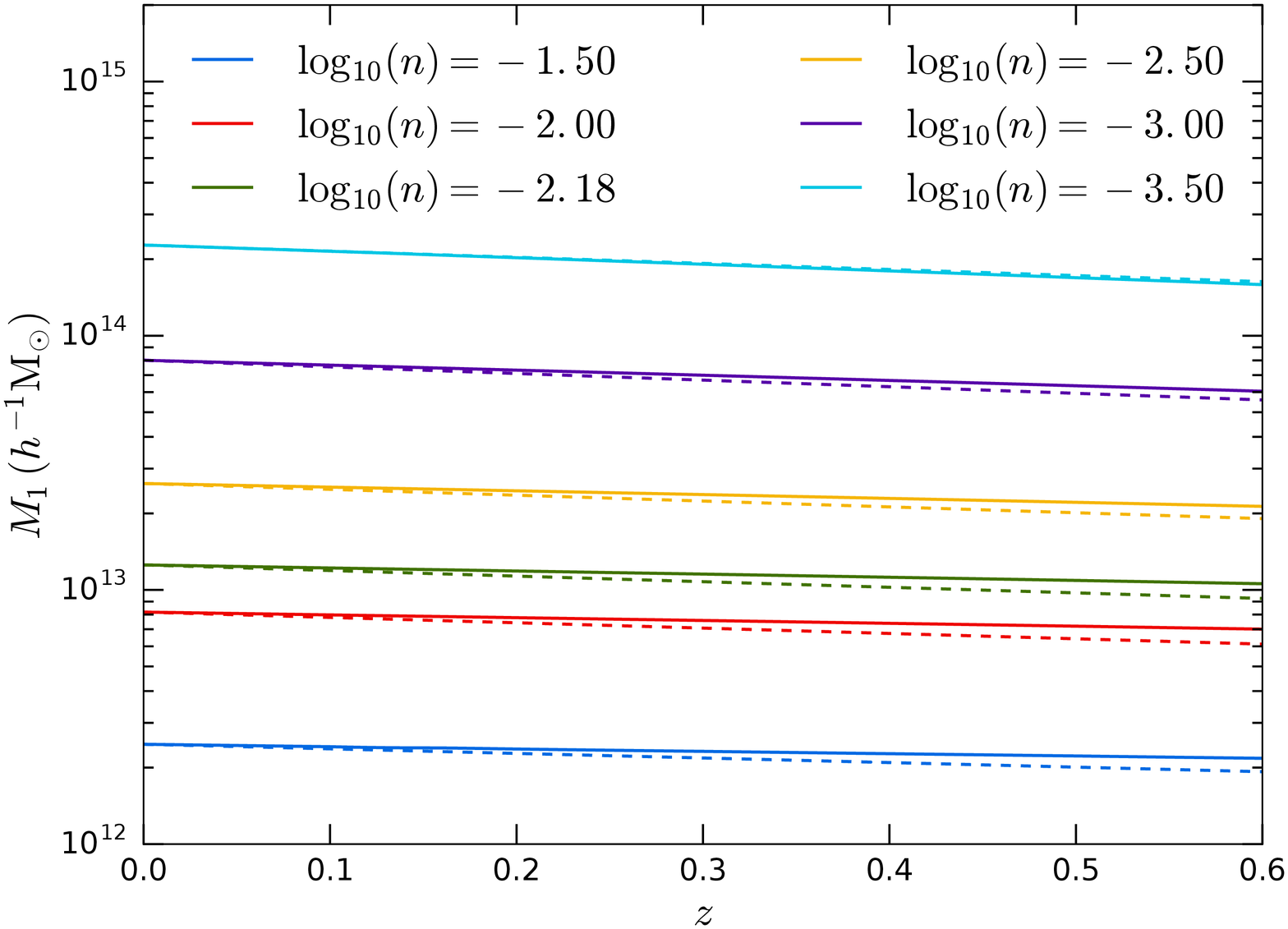}
\caption{Evolution of the HOD parameter $M_1$ with redshift for galaxy 
samples of a fixed number density, where number densities, $n$, are in
units of $h^3 \rm Mpc^{-3}$. Solid lines show the evolution in
the mock, as determined from the target luminosity function. Dashed
lines start at the same $M_1(z=0)$ as in the mock, but show the evolution found 
in \citet{Contreras2016}, as predicted
from the \citet{Gonzalez-Perez2014} version of the GALFORM 
semi-analytic galaxy formation model.}
\label{fig:M1_evolution}
\end{figure}

The evolution of the parameter $M_1$ implied by this evolution of $f$
is shown in Fig.~\ref{fig:M1_evolution} for galaxy samples of a fixed
number density, up to $z=0.6$. Since the shape of the HODs
are kept fixed for a fixed number density, but the HODs are evolved along
the mass axis, the other mass parameters $M_{\rm min}$ and $M_0$ show the
same evolution, while $\sigma_{\log M}$ and $\alpha$ are held constant. For
comparison, we also show the evolution reported in \citet{Contreras2016}
from their fit to the evolution found in the \citet{Gonzalez-Perez2014} 
version of the GALFORM semi-analytic galaxy formation model \citep{Cole2000}. 
We find that $M_1$ decreases slightly with
redshift, in remarkably close agreement with what is found in 
\citet{Contreras2016}, although the highest number density samples 
show slightly less evolution. By
construction, the ratio of the parameters $M_1/M_{\rm min}$ is kept
constant in the mock. This is in contrast to the behaviour found in 
\citet{Contreras2016},
where they reported that this ratio decreases over the same redshift range.
The amount by which this ratio decreases depends on the semi analytic
model used, and on the number density of galaxies; at most it 
decreases by $\sim 50\%$. The evolution of the HOD model could be extended
to include this change in $M_1/M_{\rm min}$, but we find that simply
keeping the mass ratio fixed produces a good match to the measured clustering
(see Fig.~\ref{fig:proj_corr_func_high_z}).

%%%%% SECTION - MOCK GALAXY CATALOGUE %%%%%

\section{Mock Galaxy Catalogue} \label{sec:mock_catalogue}

Now we describe in detail the HOD method used to populate the halo 
catalogue with galaxies, assign each galaxy a luminosity and $^{0.1}(g-r)$
colour and compare the resultant clustering in the mock with measurements
from SDSS and GAMA. These details can be skipped by the reader, but we give a brief
summary below.

Section~\ref{sec:making_gal_catalogue} describes the HOD method for populating
the halo lightcone with galaxies with luminosities. This Monte Carlo method is 
based on \citet{Skibba2006}, but extended to an evolving 5 parameter HOD. The 
number and luminosity of galaxies in each halo are randomly generated such that
the input HODs are reproduced. Central galaxies are assigned the position and 
velocity of the halo, and satellites are randomly positioned around the central,
following an NFW density profile, and assigned a random virial velocity.

The method for assigning a $^{0.1}(g-r)$ colour to each galaxy is described
in Section~\ref{sec:assigning_colours}. This is based on \citet{Skibba2009},
and randomly assigns a colour from a parametrisation of the SDSS colour
magnitude diagram. Section~\ref{sec:assigning_colours} describes
our modification to the parametrisation of the colour magnitude diagram,
which includes evolution, and is in agreement with measurements from
GAMA. The colour assigned to each galaxy depends only on its luminosity, 
its redshift, and whether it is a central or satellite galaxy; there is no 
explicit dependence on halo mass. Colour-dependent {\it k}-corrections 
derived from GAMA are described in Section~\ref{sec:k-corrections}.
The colour-dependent clustering of galaxies in the mock is shown
in Section~\ref{sec:colour_clustering}.

\subsection{Constructing the Galaxy Catalogue} \label{sec:making_gal_catalogue}

We use a modified version of the method of~\citet{Skibba2006} to populate 
the halo lightcone catalogue with galaxies, and to assign each galaxy an $r$-band 
absolute magnitude, {\it k}-corrected to $z=0.1$. \citet{Skibba2006} use a 3 parameter 
HOD in which the occupation function of central galaxies is simply a step
function; we have extended this method in order to reproduce the 5 parameter 
HOD given by Eq.~\ref{eq:hod_sat}~\&~\ref{eq:hod_cent_sk}, which adds scatter
to the luminosity of central galaxies, as required by the SDSS clustering data. 
We also use the fits to the HOD parameters 
as a function of luminosity as described in Section~\ref{sec:hods_z0.1} and
shown in Fig.~\ref{fig:hod_params}. To be consistent with the 
mass definition used in \citet{Zehavi2011}, we take the halo 
mass to be $M_{200\rm m}$; i.e. the mass enclosed by a sphere in which the average 
density is 200 times the mean density of the Universe.

For each halo, a number, $x$, is randomly drawn from the spline kernel 
probability distribution, $S(x)$ (Eq.~\ref{eq:sk_scaled}), with $\mu=0$ and 
$\sigma=1$. This introduces the scatter in the luminosity of the
central galaxy, relative to the average luminosity in a halo of this mass. 
The luminosity $L$ which is required to produce this scatter is found by solving
$x \sigma_{\log M}(L) / \sqrt{2}= \log M - \log M_{\rm min} (L)$, 
where the factor of $\sqrt{2}$ comes from how 
$\sigma_{\log M}$ is defined. Finally the central galaxy is positioned 
at the centre of the halo, with the same velocity.

\begin{figure}   % ANGULAR CLUSTERING
\includegraphics[width=\columnwidth]{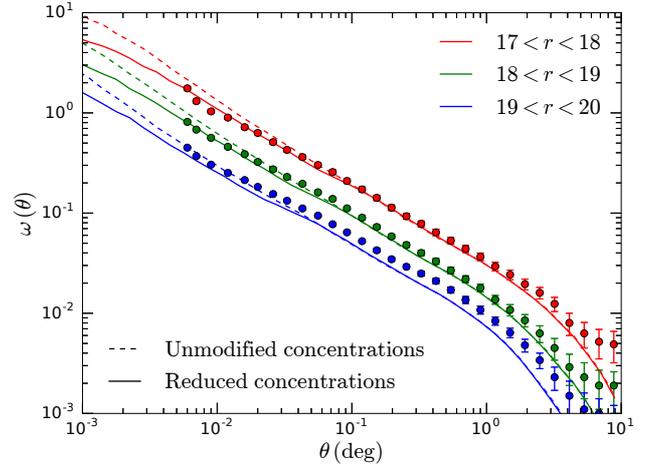}
\caption{Angular clustering of galaxies in the mock catalogue in bins of
apparent magnitude, as labelled (coloured lines). Points with error bars 
show the angular clustering of galaxies measured in the SDSS~\citep{Wang2013}.
Dashed curves show the angular clustering where satellite galaxies are 
positioned such that they follow an NFW density profile with the same,
unmodified concentration as the halo. Solid curves show the resulting
angular clustering when halo concentrations are reduced by a factor of 2.}
\label{fig:angular_clustering}
\end{figure}

To populate a halo with satellite galaxies, a minimum luminosity, $L_{\min}$, 
must first be chosen. We vary $L_{\min}$ with redshift, choosing 
it to be slightly fainter than the luminosity corresponding to $r=20$. 
This ensures that the final mock catalogue is complete to $r=20$ at 
all redshifts, while preventing galaxies that are too faint to be
observed being unnecessarily added to the catalogue.
The number of satellite galaxies to be added to each halo is drawn from a 
Poisson distribution with mean $\langle N_{\rm sat}(>L_{\rm min}|M) \rangle$, 
which is given by Eq.~\ref{eq:hod_sat}. For each satellite, a uniform random 
number $0<u<1$ is drawn, and the luminosity is found such that 
$\langle N_{\rm sat}(>L|M) \rangle / \langle N_{\rm sat}(>L_{\rm min}|M) \rangle = u$. 
The satellite galaxies are assigned a random virial velocity,
relative to the velocity of the central galaxy, which is drawn from a 
Maxwell-Boltzmann distribution with a line of sight velocity dispersion
\begin{equation}   \label{eq:los_vel_disp} % Eq - LOS velocity dispersion
\sigma^2(M) = \frac{G M_{\rm 200m}}{2R_{\rm 200m}},
\end{equation}
where $R_{\rm 200m}$ is the radius of the sphere, centred on the halo, in which
the enclosed density is 200 times the mean density of the Universe.
Finally, the satellite galaxies are positioned randomly around the centre of the halo 
such that they follow an NFW \citep{Navarro1997} density profile, which is truncated
at $R_{200\rm m}$. We find that using the same concentration, $c$,
as the halo, calculated from $c=2.16 R_{200\rm m} / R_{\rm Vmax}$, where
$R_{\rm Vmax}$ is the radius at which the maximum circular velocity
occurs, produces angular clustering which is too strong at small angular scales
compared to SDSS \citep{Wang2013}. This can be improved by reducing the 
concentration of all haloes by a factor of 2 (see Fig.~\ref{fig:angular_clustering}).
We therefore use these reduced concentrations when positioning satellite
galaxies inside each halo.

The HODs from \citet{Zehavi2011}
were fit to the projected correlation functions, using the 
mass-concentration relation of \citet{Bullock2001}, modified to be
consistent with their mass definition. This mass-concentration relation
is close to what is seen in MXXL. However, modifying the concentrations
only has a small effect on the 1-halo term of the projected correlation
functions. Down to separations of $0.1 h^{-1}\rm Mpc$, the clustering
in the mock catalogue only changes by a small amount. It is the change in
the clustering at physical scales smaller than this which causes the 
small scale angular clustering to improve, and this is below the scale at 
which the projected correlation functions were measured in SDSS. The angular 
correlation function, $\omega(\theta)$, in the mock catalogue is shown in 
Fig.~\ref{fig:angular_clustering} for
galaxies in bins of apparent magnitude, compared to the angular clustering
measured in SDSS \citep{Wang2013}. Solid lines show the 
clustering in the mock with concentrations reduced by a factor of 2, which
is in good agreement with SDSS down to a small angular separation of
20 arcsec, although for the faintest sample the clustering is a little low. 
Using unmodified concentrations results in $\omega(\theta)$
having a slope which is steeper than the SDSS measurements, shown by
the dashed curves, resulting 
in clustering which is too strong at small angular scales.
The introduction of unclustered haloes below the mass resolution has
the effect of reducing the clustering in the mock, but as we show
later in Section~\ref{sec:clustering}, this effect is small.

\subsubsection{The luminosity function of the mock}

\begin{figure}   % LUMINOSITY FUNCTION
\includegraphics[width=\columnwidth]{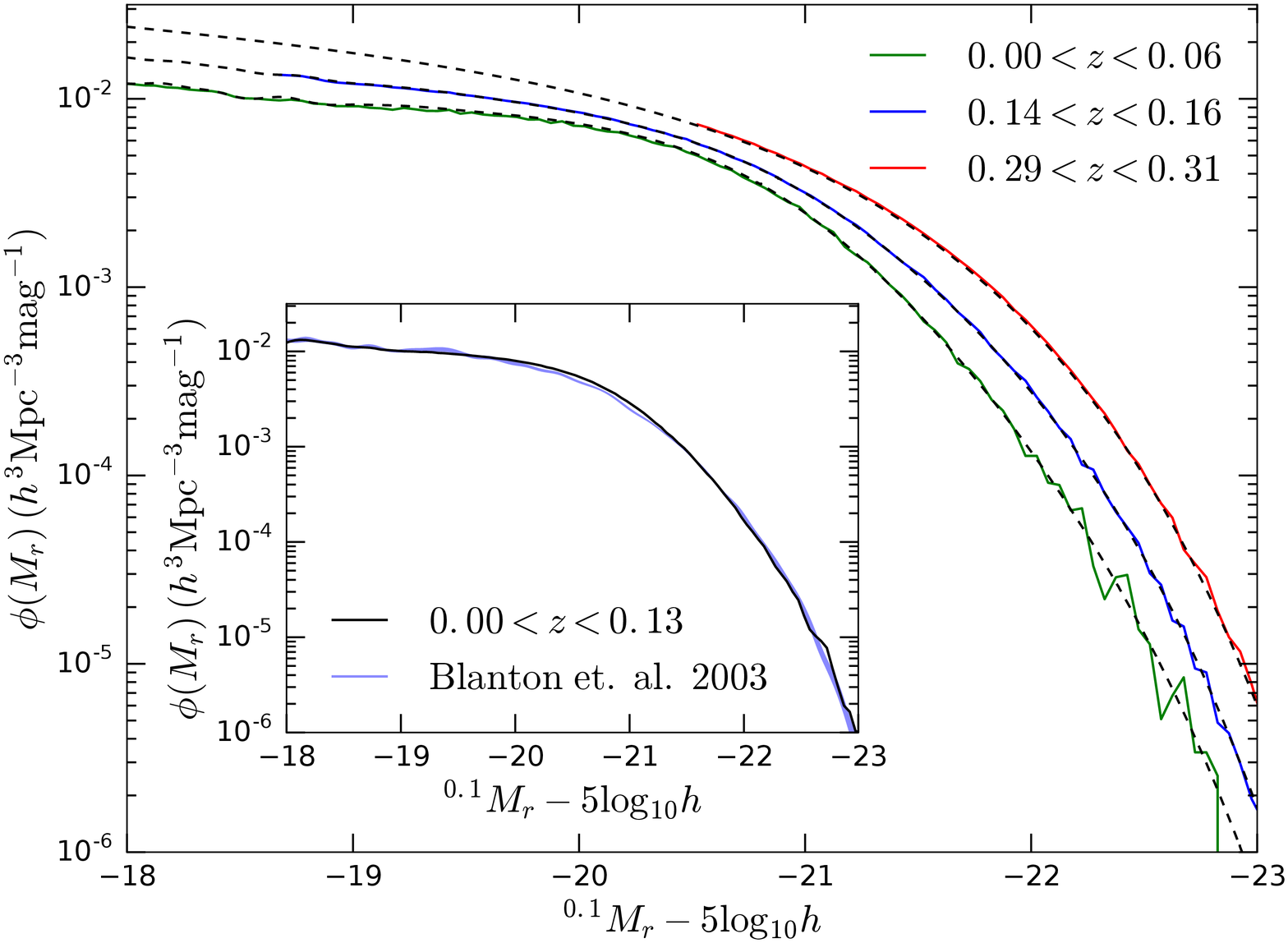}
\caption{The {\it r}-band luminosity function of galaxies in the mock catalogue 
in different
redshift bins, as indicated by the legend. Dashed lines indicate
the target luminosity function at the median redshift in each bin, which 
transitions from the SDSS luminosity function at $z<0.1$ to the GAMA 
luminosity function at $z>0.2$. The smaller panel shows the luminosity
function in the mock catalogue over the redshift range $0<z<0.13$, 
compared to the SDSS luminosity function of \citet{Blanton2003}.}
\label{fig:luminosity_function}
\end{figure}

The Petrosian {\it r}-band luminosity function of the galaxy catalogue is shown
in Fig.~\ref{fig:luminosity_function} for galaxies in three
redshift bins. The dashed lines show the target luminosity at
the median redshift of each bin, showing that this evolving
target luminosity function is reproduced in the mock catalogue.
The smaller panel in Fig.~\ref{fig:luminosity_function} compares
the luminosity function in the mock at low redshifts with the
\citet{Blanton2003} luminosity function from SDSS. Brighter than
$^{0.1}M_r=-19$, the luminosity function in the mock is in good agreement
with SDSS, which indirectly
shows that the mass function of the MXXL lightcone is close to the 
\citet{Jenkins2001} mass function assumed by \citet{Zehavi2011}, and our 
fits to the HOD parameters as a function of luminosity are a good 
approximation to the actual values. Fainter than $^{0.1}M_r=-19$, the luminosity
functions agree by construction.

\subsubsection{The redshift distribution of the mock}

\begin{figure}   % dN/dz
\includegraphics[width=\columnwidth]{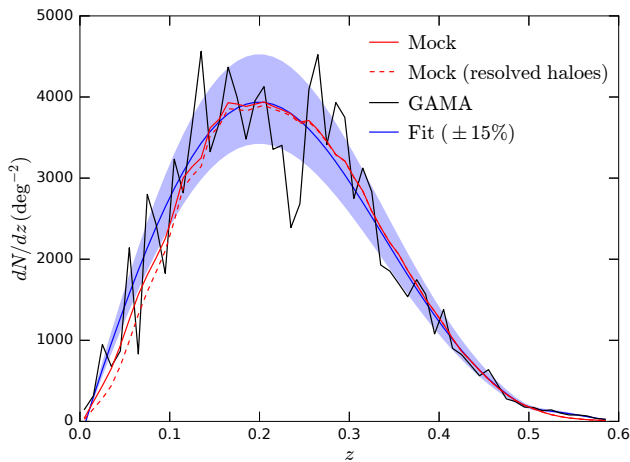}
\caption{$dN/dz$ of galaxies in the mock catalogue with $r<19.8$ (red),
compared to GAMA (black). The solid red curve shows the redshift distribution
of all galaxies, including those residing in unresolved haloes below the 
MXXL mass resolution, while the dashed red curve only includes galaxies residing
in resolved haloes. The blue curve shows a 
fit to the GAMA $dN/dz$, where the shaded region indicates $\pm 15 \%$.}
\label{fig:dN_dz}
\end{figure}

The redshift distribution of galaxies brighter than an apparent magnitude limit of 
$r=19.8$ is shown in Fig.~\ref{fig:dN_dz} (see Section~\ref{sec:k-corrections} for 
the {\it k}-corrections used), and compared to the GAMA survey. The $dN/dz$ of the mock 
catalogue is in good agreement with GAMA, within 15\% of the fitted curve at most 
redshifts. Without adding in the low mass, unresolved haloes at low redshifts (see 
Section~\ref{sec:add_unresolved_haloes}), there is a deficit in the $dN/dz$ for
$z \lesssim 0.1$ (dashed red curve); adding in these haloes increases the number of 
low redshift haloes, bringing the $dN/dz$ into better agreement with GAMA
(solid red curve).

\subsubsection{Clustering of the mock}    \label{sec:clustering}
    
\begin{figure}   % CLUSTERING LOW Z
\includegraphics[width=\columnwidth]{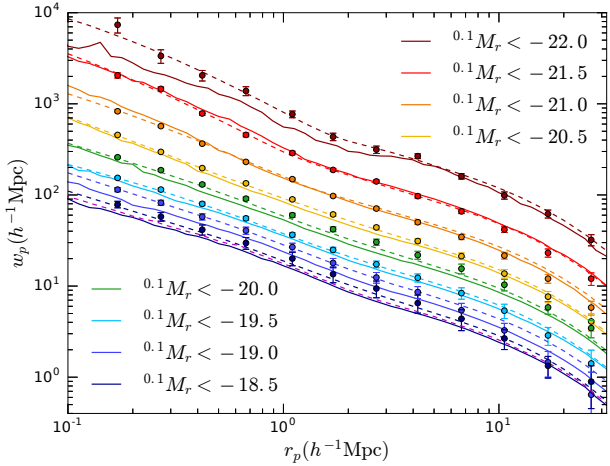}
\caption{Projected correlation functions from the galaxy catalogue (solid lines), 
compared to the projected correlation functions from SDSS \citep{Zehavi2011} 
(points with error bars) and the projected clustering predicted using the best
fitting HODs in Millennium cosmology (dashed lines), for different luminosity 
threshold samples, as indicated by the legend. For the $^{0.1}M_r < -18.5$
sample, we also show the projected clustering in the galaxy catalogue omitting 
all galaxies which reside in unresolved, unclustered haloes (magenta dashed line). 
For clarity, the results have been offset by successive intervals of 0.15 dex, 
starting at the $^{0.1}M_r < -20.5$ sample.}
\label{fig:proj_corr_func}
\end{figure}

Projected correlation functions of galaxies in the mock catalogue
are shown by the solid curves in Fig.~\ref{fig:proj_corr_func} for different luminosity 
threshold samples at $z \sim 0.1$, where we have calculated the two 
point correlation functions using the publicly available code CUTE 
\citep{Alonso2012}\footnote{\url{http://members.ift.uam-csic.es/dmonge/CUTE.html}}. 
These are compared to the measured clustering from SDSS (points with error bars), and 
the clustering predicted by the best fitting HODs (dashed lines). We use the same redshift 
ranges as the SDSS volume limited luminosity threshold samples 
\citep[see table~2 in][]{Zehavi2011}. To be consistent with the definition of magnitude
used in \citet{Zehavi2011}, magnitudes are evolved to $z=0.1$ using the evolution 
model $E(z)=Q_0 (1+Q_1(z-z_0))(z-z_0)$, where $Q_0=2$, $Q_1=-1$ and $z_0=0.1$.
The clustering in our galaxy catalogue 
is in reasonable agreement with the projected correlation functions
measured from SDSS. 

The small differences in the large scale clustering between the mock catalogue
and the clustering predicted by the best fitting HODs can be understood by 
comparing the HODs in Fig.~\ref{fig:HODs}. For example, the $^{0.1}M_r<-19$ 
sample is slightly less clustered in the mock. The fit to the HOD has a smaller
$M_{\rm min}$ than the best fitting SDSS HOD, meaning that this sample contains 
more low mass haloes. These haloes are less biased, and therefore the 
clustering is reduced compared to SDSS. The $^{0.1}M_r<-22$
sample contains more high mass haloes, and should therefore be more clustered
than SDSS, but the opposite is seen. This is because the brightest 
samples cover a wider redshift range, and are affected more by the 
evolution of the HODs.

The clustering of galaxies in the mock catalogue is also affected by
the introduction of haloes below the MXXL mass resolution, which are unclustered. 
Adding these haloes will therefore have the effect of reducing the measured 
galaxy clustering. The galaxies which
reside in these haloes are faint, and have low redshifts, and so the faintest
galaxy samples in Fig.~\ref{fig:HODs} are affected by this more than the bright 
samples. For the $^{0.1}M_r < -18.5$ sample, we illustrate the size of this effect: 
the magenta dashed curve shows the projected correlation function with galaxies
residing in unresolved haloes omitted. Including these galaxies reduces the
clustering, but only by a very small amount.

We have checked that if we modify our fits to the HOD parameters to agree exactly
with the best fitting SDSS parameters at one magnitude, and do not evolve the
HODs, we reproduce the SDSS correlation functions very closely for that
magnitude limit.

\begin{figure*}   %CLUSTERING HIGH Z
\includegraphics[width=0.75\textwidth]{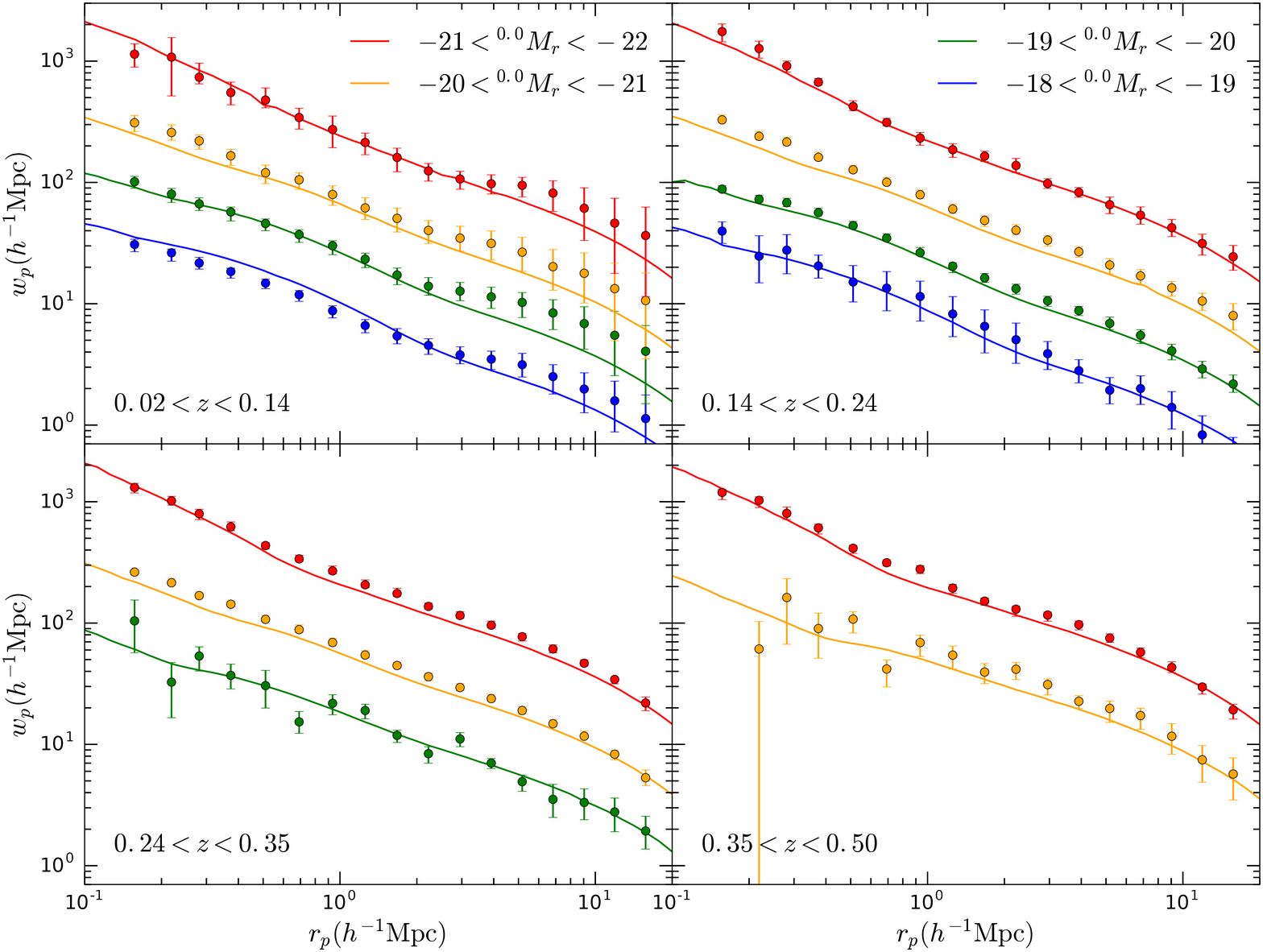}
\caption{Projected correlation functions in different redshift bins for
galaxies in the mock catalogue (lines). Points with error bars show the 
clustering of galaxies from GAMA~\citep{Farrow2015}. Different colours
indicate bins in $^{0.0}M_r$ absolute magnitude. Lines are offset by
0.4 dex relative to the $-21 < ^{0.0}M_r < -20$ samples, for clarity.}
\label{fig:proj_corr_func_high_z}
\end{figure*}

In Fig.~\ref{fig:proj_corr_func_high_z} we show the projected correlation
functions in the mock catalogue at high redshifts, compared to the
clustering measured in GAMA by \citet{Farrow2015}, which has a high
completeness of galaxy pairs \citep{Robotham2010}. Here, the magnitude
ranges are defined for magnitudes {\it k}-corrected to a reference redshift
of $z=0$, denoted as $^{0.0}M_r$, which have also had evolutionary corrections 
applied. To
be consistent with \citet{Farrow2015}, we use the same evolutionary
correction $E(z)=-Q(z-z_{\rm ref})=-1.45z$ for this comparison. We find that 
the clustering
at high redshifts is in good agreement with the clustering seen in GAMA.
The agreement is least good in the lowest redshift bin, but at low 
redshifts we find good agreement with SDSS, which covers a much larger
area of the sky than GAMA.

\subsection{Assigning Colours} \label{sec:assigning_colours}

We use the method of \citet{Skibba2009} to assign each galaxy a 
$^{0.1}(g-r)$ colour, where $g$ and $r$ are SDSS DR7
model magnitudes \citep{Abazajian2009,Baldry2010}. 
This method parametrises the red and blue sequence 
of the colour-magnitude diagram as two Gaussians with a mean and rms that 
are linear functions of magnitude. A galaxy is randomly chosen to be red 
or blue, then a colour is drawn from the appropriate Gaussian. We have
modified the parametrisation of the colour-magnitude diagram given in
\citet{Skibba2009} to bring the faint end into agreement with the
colour-magnitude diagram from GAMA, and to add evolution. For those 
interested, this is described in detail below. For clarity, the 0.1
superscript has been omitted from the equations in the following 
subsections.

\subsubsection{Low redshift} \label{sec:colours_low_z}

For redshifts $z<0.1$, we use the parametrisation used by~\citet{Skibba2009}
that produces a good approximation to the SDSS colour-magnitude diagram,
which we summarise below. However, we make some slight modifications
to the parametrisation to also bring it into agreement with GAMA at
faint magnitudes.

The mean and rms of the red and blue sequences are given by
\begin{eqnarray} \label{eq:red_seq} % Eq - red sequence
\nonumber
\langle g-r | M_r \rangle_{\rm red}^{\rm Skibba} &=& 0.932 - 0.032(M_r + 20) \\
\rms (g-r | M_r)_{\rm red}^{\rm Skibba} &=& 0.07 + 0.01(M_r + 20),
\end{eqnarray}
and
\begin{eqnarray} \label{eq:blue_seq} % Eq - blue sequence
\nonumber 
\langle g-r | M_r \rangle_{\rm blue}^{\rm Skibba} &=& 0.62 - 0.11(M_r + 20) \\
\rms (g-r | M_r)_{\rm blue}^{\rm Skibba} &=& 0.12 + 0.02(M_r + 20).
\end{eqnarray}
The total fraction of galaxies which are blue is also parametrised 
as a linear function of magnitude, given by
\begin{equation} \label{eq:f_blue_tot} % Eq - total blue fraction
f_{\rm blue}^{\rm Skibba}(M_r) = 0.46 + 0.07(M_r + 20).
\end{equation}

These relations from \citet{Skibba2009} produce a colour-magnitude
diagram which is in good agreement with SDSS at the bright
end. However, the faint end does not agree with what is seen
in GAMA \citep[e.g. the first panel in figure~6 of][]{Loveday2012}.
Firstly, at $^{0.1}M_r = -16$, all the galaxies should lie on
the blue sequence, while the fraction of galaxies which are 
blue given by Eq.~\ref{eq:f_blue_tot} is 0.74. At faint magnitudes,
we instead use a blue fraction given by
\begin{equation}
f_{\rm blue}^{\rm faint}(M_r) = 0.4 + 0.2(M_r + 20),
\end{equation}
so the total fraction of blue galaxies is
\begin{equation} \label{eq:f_blue_tot_new}
f_{\rm blue}(M_r) = {\rm max}\{f_{\rm blue}^{\rm faint}(M_r), f_{\rm blue}^{\rm Skibba}(M_r)\};
\end{equation}
$f_{\rm blue}(M_r)$ is capped so it is always in the range $0 \leq f_{\rm blue}(M_r) \leq 1$.
Another issue with the parametrisation of \citet{Skibba2009} is that
faint galaxies which lie on the blue sequence are too blue in comparison 
to the galaxies in GAMA. At $^{0.1}M_r=-18.7$, we transition to
a flatter blue sequence, given by
\begin{equation}
\langle g-r | M_r \rangle_{\rm blue}^{\rm faint} = 0.4 - 0.03(M_r + 16).
\end{equation}

If the fraction of satellite galaxies that are blue, $f_{\rm sat}^{\rm blue}(M_r)$, is specified, then
the mean colour of satellite galaxies is given by
\begin{eqnarray} \label{eq:av_sat_col}
\langle g-r | M_r \rangle_{\rm sat} &=& f_{\rm sat}^{\rm blue}(M_r)\langle g-r | M_r \rangle_{\rm blue} \\
\nonumber &+& (1-f_{\rm sat}^{\rm blue}(M_r))\langle g-r | M_r \rangle_{\rm red}.
\end{eqnarray}
Conversely, Eq.~\ref{eq:av_sat_col} can be rearranged, and the mean satellite colour can be used to specify
the fraction of satellites that are blue,
\begin{equation} \label{eq:f_blue_sat}    % Eq - probability red satellite
f_{\rm sat}^{\rm blue}(M_r) = \frac{\langle g-r | M_r \rangle_{\rm sat} - \langle g-r | M_r \rangle_{\rm red}}{\langle g-r | M_r \rangle_{\rm blue} - \langle g-r | M_r \rangle_{\rm red}},
\end{equation}
\citep[equation~8 from][but for blue galaxies]{Skibba2009}.
The average colour of a satellite galaxy is parametrised by
\citet{Skibba2009} as
\begin{equation} \label{eq:sat_col_Skibba}    % Eq - Mean satellite colour Skibba
\langle g-r | M_r \rangle_{\rm sat}^{\rm Skibba} = 0.83 - 0.08(M_r + 20).
\end{equation}
Modifying the mean satellite colour has the effect of changing the
strength of the colour dependent clustering. We find that we get
a better agreement with the clustering in SDSS by modifying the
mean satellite colour to
\begin{equation} \label{eq:sat_col_mod}    % Eq - Mean satellite colour modified
\langle g-r | M_r \rangle_{\rm sat} = 0.86 - 0.065 (M_r + 20).
\end{equation}

At, for example, $^{0.1}M_r=-16$, the fraction of blue satellites given by 
Eq.~\ref{eq:f_blue_sat}~\&~\ref{eq:sat_col_mod} is less than 1, meaning that
some satellite galaxies are red. However, all galaxies at this magnitude
should lie on the blue sequence, as determined from Eq~\ref{eq:f_blue_tot_new}.
In order to achieve the correct $f_{\rm blue}$ from Eq~\ref{eq:f_blue_tot_new},
the fraction of satellites which are blue must be 
$(f_{\rm blue} - f_{\rm cen}) / f_{\rm sat}$ if all central galaxies are on 
the blue sequence. If the value of $f_{\rm sat}^{\rm blue}$ 
is greater than this, it is still possible to get the correct $f_{\rm blue}$
by making central galaxies red, but $f_{\rm sat}^{\rm blue}$ cannot be less than
this. To ensure that at faint magnitudes we get the total fraction of blue
galaxies given by Eq~\ref{eq:f_blue_tot_new}, we take the fraction of blue
satellites to be
\begin{equation} \label{eq:f_blue_sat_new}
f_{\rm sat}^{\rm blue}(M_r) = {\rm max}\left\{f_{\rm sat}^{\rm blue}(M_r), \frac{f_{\rm blue}(M_r)-f_{\rm cen}(M_r)}{f_{\rm sat}(M_r)}\right\}.
\end{equation}

The fraction of central galaxies that are blue can then be determined
from $f_{\rm blue}(M_r)$ and $f_{\rm sat}^{\rm blue}(M_r)$. However, \citet{Skibba2009}
erroneously state that the fraction of central galaxies which are blue is
\begin{equation} \label{eq:f_cent_blue_wrong} % Eq - fraction blue centrals (wrong)
f_{\rm cen}^{\rm blue}(M_r) = f_{\rm blue}(M_r) / f_{\rm cen}(M_r),
\end{equation}
where $f_{\rm cen}(M_r)$ is the fraction of galaxies which are centrals. 
Eq.~\ref{eq:f_cent_blue_wrong} is only true if all satellite galaxies are red; 
since a significant fraction of faint satellites are blue, the fraction of blue 
central galaxies needs to be reduced to ensure we get the correct total 
fraction of blue galaxies given by Eq.~\ref{eq:f_blue_tot}. This is achieved 
by changing Eq.~\ref{eq:f_cent_blue_wrong} to
\begin{equation} \label{eq:f_cent_blue_correct}    % Eq - fraction blue centrals (corrected)
f_{\rm cen}^{\rm blue}(M_r) = \frac{f_{\rm blue}(M_r) - f_{\rm sat}^{\rm blue}(M_r)(1-f_{\rm cen}(M_r))}{f_{\rm cen}(M_r)}.
\end{equation}

For each galaxy, a uniform random number $x$ is drawn in the interval $0<x<1$. 
For central galaxies, if $x < f_{\rm cen}^{\rm blue}(M_r)$ (given by 
Eq.~\ref{eq:f_cent_blue_correct}), the galaxy is blue, and a colour is drawn randomly 
from the Gaussian distribution defined by Eq.~\ref{eq:blue_seq}, otherwise it is red,
and the colour is drawn from Eq.~\ref{eq:red_seq}. Similarly, satellite galaxies are
assigned to the blue sequence if $x < f_{\rm sat}^{\rm blue}(M_r)$, and the red
sequence otherwise.

\subsubsection{Evolution of colours with redshift}
 
The colour magnitude diagram evolves with redshift, as seen for example 
in figure~6 of \citet{Loveday2012} from GAMA. We therefore
need to evolve the expressions given in Section~\ref{sec:colours_low_z}
in order to produce a mock which has a realistic distribution of colours
at these redshifts. In the GAMA data at high redshifts, only the brightest 
tip of the red and blue sequences can be seen, making it difficult to
constrain their slopes. We therefore keep the slope of the red and
blue sequence fixed with redshift.

We keep the red and blue sequences fixed at $z<0.1$, and evolve them
with redshift as
\begin{eqnarray}     % Eq - red sequence evolution
\nonumber \langle g-r | M_r \rangle_{\rm red}(z) &=& \langle g-r | M_r \rangle_{\rm red} \\
\nonumber &-& 0.18({\rm min}\{z,0.4\}-0.1) \\ 
\rms (g-r | M_r)_{\rm red}(z) &=& \rms (g-r | M_r)_{\rm red} \\
\nonumber &+& 0.5(z-0.1) + 0.1(z-0.1)^2
\end{eqnarray}
and
\begin{eqnarray}     % Eq - blue sequence evolution
\nonumber \langle g-r | M_r \rangle_{\rm blue}(z) &=& \langle g-r | M_r \rangle_{\rm blue} \\
\nonumber &-& 0.25({\rm min}\{z,0.4\}-0.1) \\
\nonumber \rms (g-r | M_r)_{\rm blue}(z) &=& \rms (g-r | M_r)_{\rm blue}  \\
&+& 0.2(z-0.1),
\end{eqnarray}
respectively, where we stop evolving the mean of the sequences above $z=0.4$ in order 
to prevent too many high redshift galaxies being assigned as blue.

The mean satellite colour is also evolved as,    
\begin{equation} % Eq - satellite colour evolution
\langle g-r | M_r \rangle_{\rm sat}(z) = \langle g-r | M_r \rangle_{\rm sat} - 0.18 (z-0.1),
\end{equation}
and the fraction of blue galaxies is evolved as
\begin{equation}
f_{\rm blue}(M_r)(z) = 0.2M_r + 4.4 + 1.2(z-0.1) + 0.5(z-0.1)^2.
\end{equation}

Fig.~\ref{fig:colour_dist} shows the distribution of colours in the mock
catalogue compared to GAMA for galaxies in different redshift and
magnitude bins. Our parametrisation of the colour evolution is able
to produce a good approximation to the GAMA colour distributions 
at all redshifts. 

To evolve the luminosity function, we have assumed
a fixed $Q$ parameter for all galaxies. We note that 
\citet{Loveday2012,Loveday2015} hint that red and blue galaxies 
evolve differently, with a different $Q_{\rm red}$ and $Q_{\rm blue}$. 
However, the assumption of fixed $Q$ with this parametrisation
is able to reproduce the observed colour-magnitude diagram.

\subsection{Colour dependent k-corrections}  \label{sec:k-corrections}

In the mock catalogue, we use the HOD method to assign each galaxy an {\it r}-band
absolute magnitude $^{0.1}M_r$, and the method outlined above to randomly
generate a $^{0.1}(g-r)$ colour. However, the apparent magnitude, $r$, is the 
quantity which would be measured directly by the survey, and this is related to the
absolute magnitude, $^{0.1}M_r$, through the equation
\begin{equation}     % Eq - absolute magnitude
^{0.1} \! M_r - 5\log_{10}h =  r - 5\log_{10}(d_L(z)) - 25 - ^{0.1} \!\! k(z),
\end{equation}
where $d_L(z)$ is the luminosity distance in units of $h^{-1}\rm{Mpc}$, 
and $^{0.1}k(z)$ is the {\it k}-correction. The superscript $0.1$ denotes that the 
magnitude has been {\it k}-corrected to a reference redshift of $z_{\rm ref}=0.1$. 
In order to calculate an apparent magnitude for each galaxy in the mock catalogue,
we use colour-dependent {\it k}-corrections derived from the GAMA survey, similar 
to those given in table~1 of \citet{McNaught-Roberts2014}, except for 
{\it k}-correcting to $z_{\rm ref}=0.1$, rather than $z_{\rm ref}=0$.
    
The {\it k}-correction for each individual galaxy in GAMA is fit with a 4th order 
polynomial of the form
\begin{equation} \label{eq:k_correction} % Eq - k-correction polynomial
^{0.1}k(z) = \sum^4_{i=0}A_i(z-0.1)^{4-i}.
\end{equation}
The median {\it k}-correction is then found in 7 equally spaced bins of $^{0.1}(g-r)$ 
colour. Strictly speaking, the constant term in Eq.~\ref{eq:k_correction} should 
have the value $A_4=-2.5\log_{10}(1+z_{\rm ref})$ \citep{Hogg2002};
\citet{McNaught-Roberts2014} do not require this, but they end up with values of 
$A_4$ close to $0$ at their $z_{\rm ref}=0$. We force our {\it k}-corrections to 
have the value $A_4=-2.5\log_{10}(1.1)$ at our $z_{\rm ref}=0.1$, but this 
only has a small effect on the {\it k}-corrections.

Using 7 distinct {\it k}-corrections based on colour leads to artificial
features being added to the mock catalogue; for example step features can
be seen in the colour-magnitude diagram at the boundaries between colour
bins. In order to remove these features, we interpolate the {\it k}-corrections
between the median colour in each bin.

\begin{figure}   % K-CORRECTIONS
\includegraphics[width=\columnwidth]{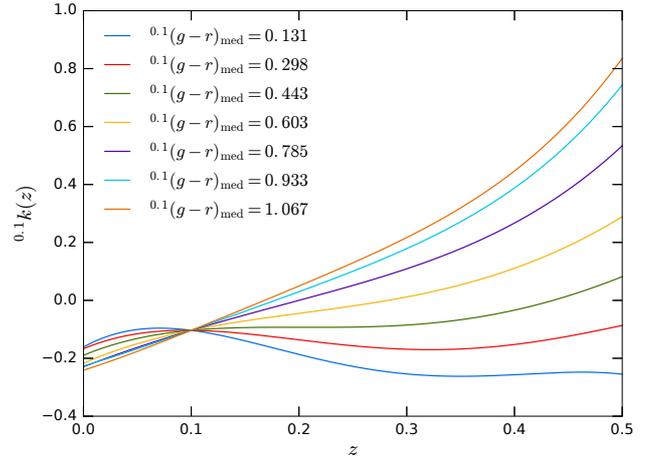}
\caption{Median $^{0.1}(g-r)$ colour-dependent {\it k}-correction for galaxies in
GAMA as a function of redshift, in 7 equally spaced bins of colour. The 
colour of each line indicates the colour bin; the median colour is indicated
in the legend.}
\label{fig:k_corrections}
\end{figure}

\begin{figure}   % COLOUR DISTRIBUTIONS AT DIFFERENT REDSHIFTS
\includegraphics[width=\columnwidth]{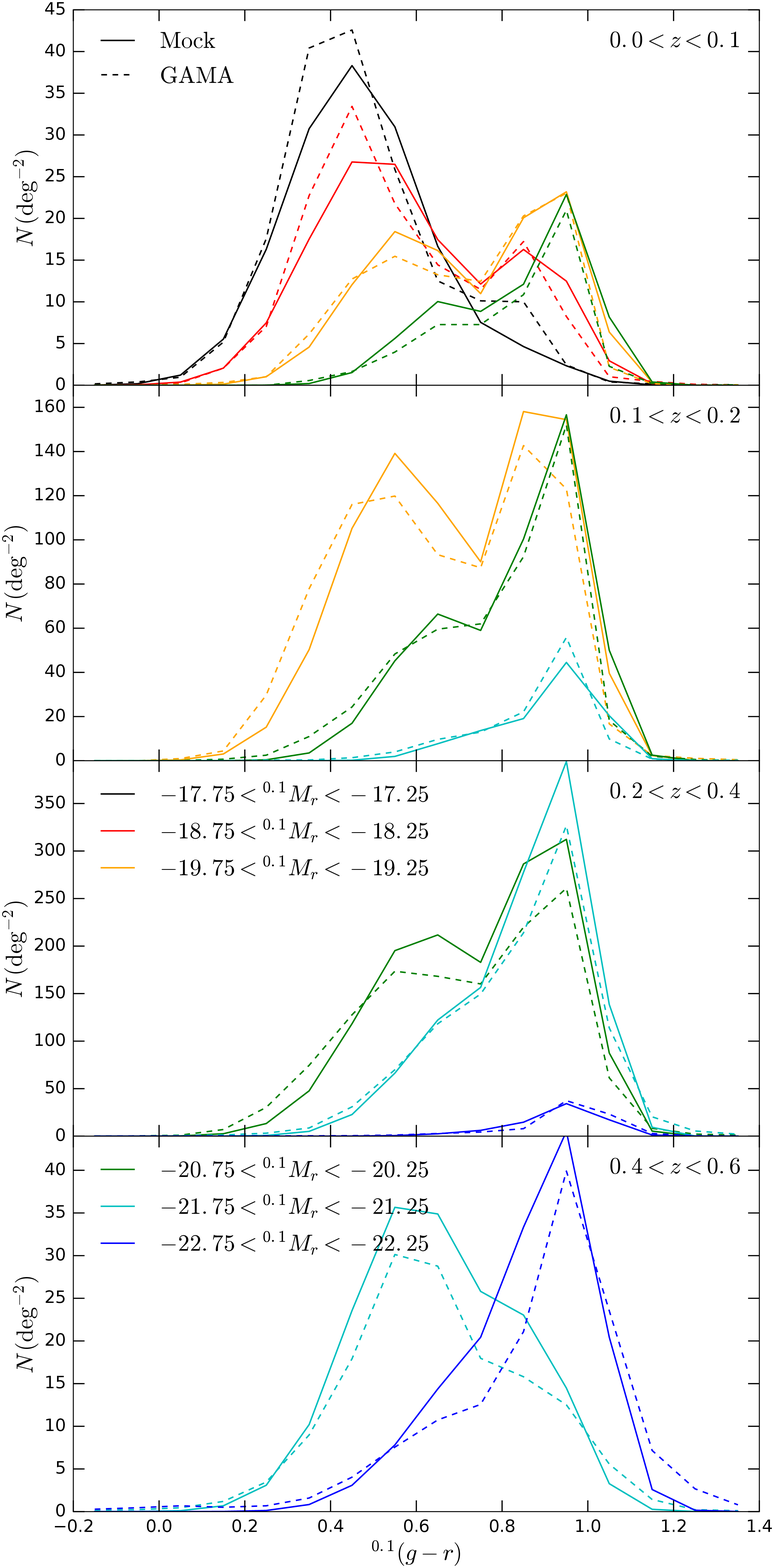}
\caption{Distribution of $^{0.1}(g-r)$ colours in the mock catalogue (solid lines) 
compared to GAMA (dashed lines). Each panel shows the colour distributions of
galaxies in a certain redshift range. Different ranges in absolute magnitude
are indicated by the colour of the line, as shown in the legend, which is 
split over several panels.}
\label{fig:colour_dist}
\end{figure}

Fig.~\ref{fig:k_corrections} shows the polynomial fits to the {\it k}-corrections
as a function of redshift. By definition, all the curves cross at 
$^{0.1}k(z=0.1)=-2.5\log_{10}(1.1)\approx -0.103$. The polynomial coefficients
are shown in Table~\ref{tab:k_corrections}.

\begin{table}   % TABLE - K-CORRECTION COEFFICIENTS
\caption{Polynomial coefficients of the median {\it k}-corrections of galaxies
in GAMA in equally spaced bins of $^{0.1}(g-r)$ colour, as defined in
Eq.~\ref{eq:k_correction}. $^{0.1}(g-r)_{\rm med}$ is the median colour
in each bin, and $A_i$ are the polynomial coefficients. The constant
term $A_4=-2.5\log_{10}(1.1)\approx -0.103$, as described in the text.}
\label{tab:k_corrections}
\begin{tabular}{ccccc}
\hline
$^{0.1}(g-r)_{\rm med}$ & $A_0$ & $A_1$ & $A_2$ & $A_3$ \\
\hline
0.131 & -45.33 & 35.28  & -6.604  & -0.4805  \\
0.298 & -20.08 & 20.14  & -4.620  & -0.04824 \\
0.443 & -10.98 & 14.36  & -3.676  &  0.3395  \\
0.603 & -3.428 & 9.478  & -2.703  &  0.7646  \\
0.785 &  6.717 & 3.250  & -1.176  &  1.113   \\
0.933 & 16.76  & -2.514 &  0.3513 &  1.307   \\
1.067 & 20.30  & -4.189 &  0.5619 &  1.494   \\
\hline
\end{tabular}
\end{table}

\subsection{Colour dependent clustering in the mock} \label{sec:colour_clustering}

\begin{figure}   % CLUSTERING RED/BLUE LOW Z
\includegraphics[width=\columnwidth]{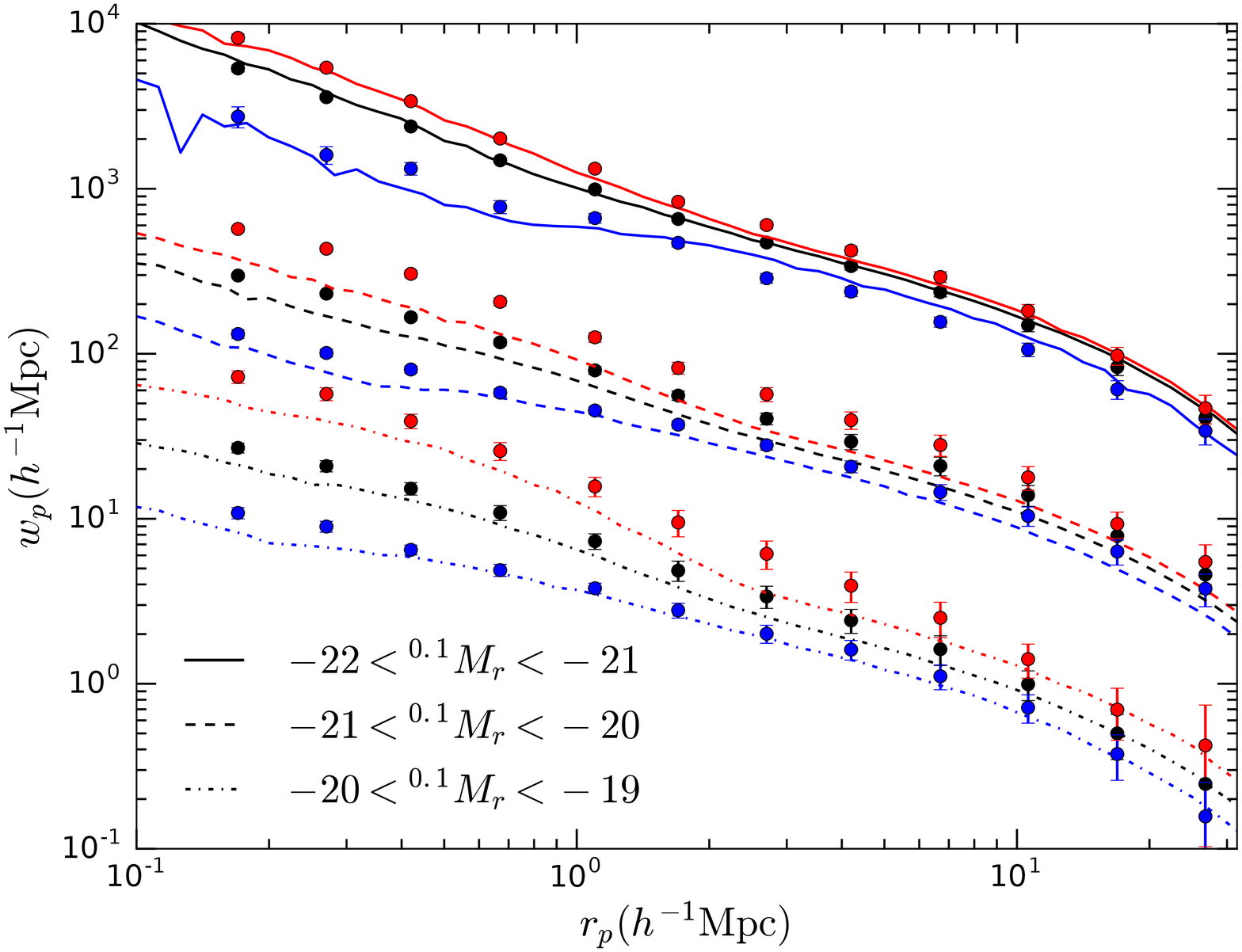}
\caption{Projected correlation functions of red and blue galaxy samples in the 
mock catalogue at low redshifts (lines) compared to the SDSS volume limited
samples \citep{Zehavi2011} (points with error bars) for different magnitude 
bins. The clustering of all
galaxies in a sample is shown in black, while clustering for red and blue
galaxies, defined by the colour cut $^{0.1}(g-r)_{\rm cut}=0.21-0.03 \, ^{0.1}M_r$, 
is shown in red and blue, respectively. Line style indicates the
magnitude bin, as shown by the legend. For clarity, magnitude samples are successively 
offset by 1 dex from the $-21 < ^{0.1}M_r < -20$ samples.}
\label{fig:proj_corr_func_colours_low_z}
\end{figure}

The projected correlation function of galaxies in the mock at low redshifts, 
split by red and blue galaxies, is shown in 
Fig.~\ref{fig:proj_corr_func_colours_low_z} for different bins in
absolute magnitude and compared to the clustering in the corresponding
volume limited samples from SDSS \citep{Zehavi2011}, where the red
and blue samples are defined using the same colour cut as their equation~13. 
In the 
SDSS data, red galaxies are clustered more strongly than blue galaxies,
since red elliptical galaxies are more likely to reside in more massive 
haloes, which are more strongly biased \citep{Eisenstein2005}. As the samples 
get fainter, the strength of the colour dependence becomes stronger. These
trends are reproduced in the mock catalogue, using the modified
satellite colour in Eq.~\ref{eq:sat_col_mod}.

\begin{figure}   % CLUSTERING RED/BLUE HIGH Z
\includegraphics[width=\columnwidth]{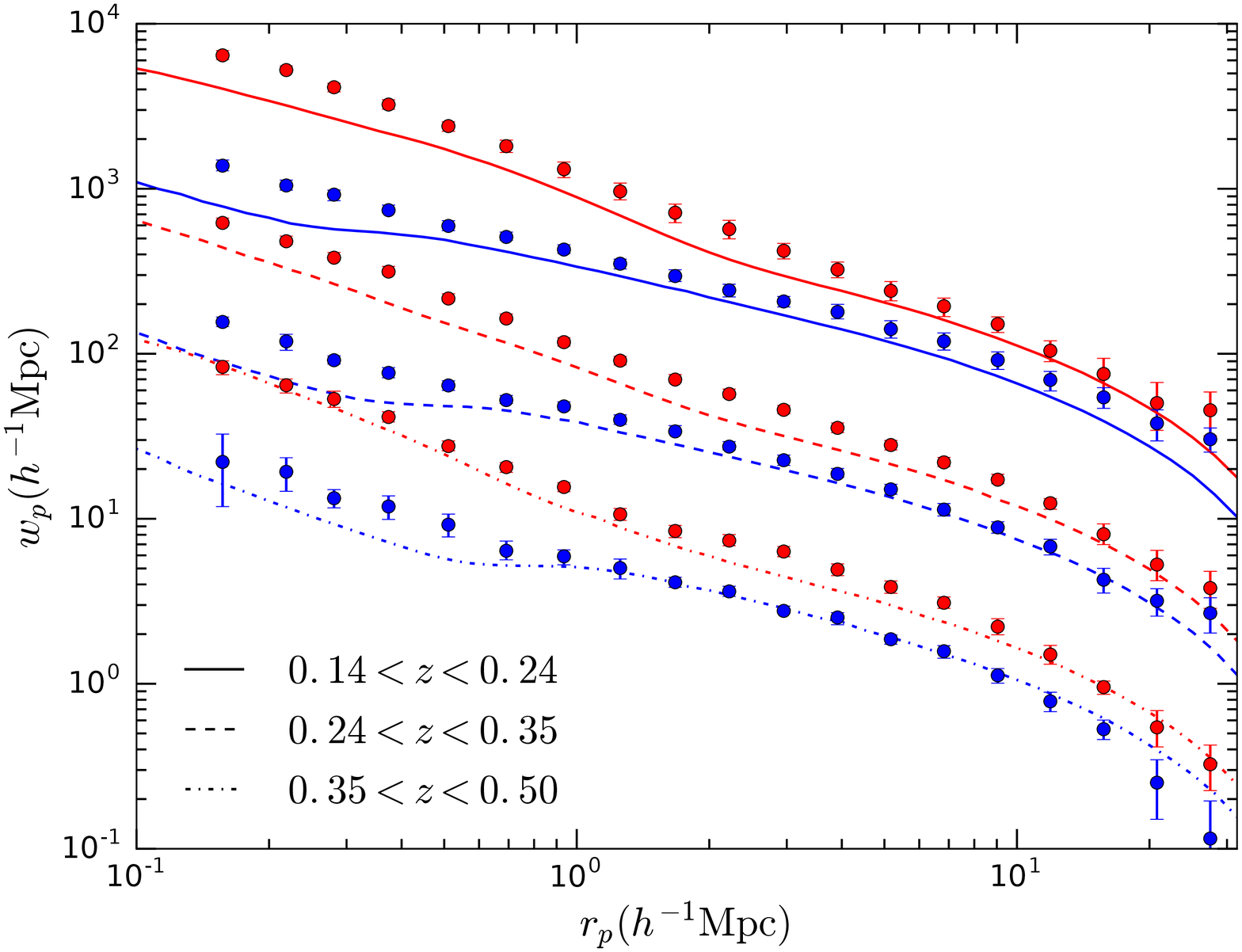}
\caption{Projected correlation functions of red and blue galaxy samples
at high redshift in the mock catalogue (lines), compared to GAMA (points
with error bars) \citep{Farrow2015}. Red and blue lines indicate red and
blue galaxy samples, where the colour cut is defined as 
$^{0.0}(g-r)_{\rm cut}=-0.03(^{0.0}M_r + 20.6) + 0.678$. For each sample of 
galaxies in the mock, the same $^{0.0}M_r$ magnitude range is used as the 
GAMA galaxy sample. The style of the line indicates the redshift range, 
as shown in the legend. Redshift samples are successively offset from the 
$0.24<z<0.35$ samples by 1 dex for clarity.}
\label{fig:proj_corr_func_colours_high_z}
\end{figure}

Projected correlation functions for red and blue galaxies are also shown
for higher redshift galaxies in Fig.~\ref{fig:proj_corr_func_colours_high_z},
compared with the clustering seen in GAMA \citep{Farrow2015}. The galaxy
samples are defined using the same $^{0.0}M_r$ magnitude ranges as figure~14
of \citet{Farrow2015}, and using the same $^{0.0}(g-r)$ colour cut (their
equation~4), where the superscript 0.0 denotes that these magnitudes
are {\it k}-corrected to a reference redshift of $z_{\rm ref}=0$. 
The clustering of the red and blue galaxies in the mock
is in reasonable agreement with the GAMA data. 

\section{Applications} \label{sec:applications}

As shown in Section~\ref{sec:mock_catalogue}, the galaxies in the mock
catalogue have realistic clustering, which is in agreement with measurements
from SDSS and GAMA. The galaxies also have a realistic distribution
of $^{0.1}(g-r)$ colours at different redshifts. Future surveys,
such as DESI and Euclid, aim to make measurements of the BAO and redshift
space distortions, which probe larger scales than have been considered
so far. Here, we show as an example some of the
measurements that can be made using this mock catalogue.

\subsection{BAO}

\begin{figure}   % GALAXY CLUSTERING -- BAO
\includegraphics[width=\columnwidth]{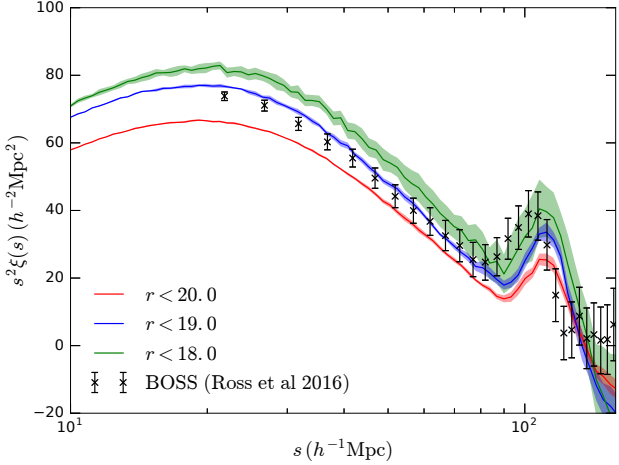}
\caption{Large scale redshift-space correlation function in the galaxy
catalogue, scaled by $s^2$, for different apparent magnitude threshold samples, 
as indicated
by the colour. For each sample, the solid curve is the clustering calculated
over the full sky, and the shaded area is the error on the mean of four quadrants.
Crosses with error bars indicate the measured clustering from BOSS in 
the redshift range $0.2<z<0.5$ \citep{Ross2016}, divided by a factor of 1.5,
to make the comparison easier.}
\label{fig:bao_galaxy}
\end{figure}

As described in Section~\ref{sec:mxxl} and shown in Fig.~\ref{fig:xi_bao}, the 
large box size of the MXXL simulation enables the BAO feature to be seen 
clearly in the clustering of haloes. Here, we show that the BAO can also
be seen in measurements of the redshift-space galaxy clustering. 
Fig.~\ref{fig:bao_galaxy}
shows the large scale redshift-space correlation function for several
apparent magnitude threshold galaxy samples, using a redshift weighting
$w(z)=1 / (1+ 4\pi J_3 \bar{n}(z))$ \citep{Efstathiou1990}, where $\bar{n}(z)$
is the number density of galaxies in the sample at redshift $z$, and 
$J_3=\int \xi r^2 dr$, where we have assumed $4\pi J_3=3 \times 10^4 h^3 \rm Mpc^{-3}$. 
The BAO peak can be seen in all samples, but the errors in the correlation function 
are largest for the $r<18.0$ sample. The $r<20.0$ sample contains fainter galaxies, 
and covers a larger volume, which greatly reduces the errors.
For comparison, the crosses with error bars show measurements of the 
BAO from the Baryon Oscillation Spectroscopic Survey (BOSS) \citep{Reid2016}
for galaxies in the redshift range $0.2<z<0.5$ \citep{Ross2016}.
The BAO scale in the mock catalogue is $\sim 7 \%$ larger than is 
measured in BOSS. This is consistent with the difference in cosmology
between that used in the MXXL simulation and the best fit to observations,
including the BOSS results, and
is mostly driven by the difference between $\Omega_{\rm m}=0.25$ in MXXL
and $\Omega_{\rm m}=0.31$ in the Planck cosmology.
The amplitude of the BAO peak in the mock catalogue also differs with the BOSS
results, but we have not made any attempt to match the BOSS
colour selection.

\subsection{Redshift Space Distortions}

The two-point correlation function $\xi(s,\mu)$, in bins of $s$ 
and $\mu$, can be decomposed into multipoles \citep{Hamilton1992}
\begin{equation}
\xi(s, \mu) = \sum_l \xi_l(s)P_l(\mu),
\end{equation}
where $s$ is the separation between a pair of galaxies in redshift space,
$\mu=\cos \theta$ is the cosine of the angle between the vector $\mathbf{s}$
and the line of sight, and $P_l(\mu)$ is the {\it l}\textsuperscript{th} order Legendre 
polynomial. The multipoles can be determined from the measured $\xi(s,\mu)$ by
evaluating the integral
\begin{equation}
\xi_l(s) = \frac{2l+1}{2} \int^1_{-1} \xi(s,\mu)P_l(\mu)d\mu.
\end{equation}
Due to the symmetry $\xi(s,\mu)=\xi(s,-\mu)$, all odd-numbered terms are zero, and
in linear theory, it is only the monopole, $\xi_0(s)$, quadrupole, $\xi_2(s)$, and
hexadecapole, $\xi_4(s)$, that are non-zero. The amplitude of these multipoles
depends on the strength of the redshift space distortions, and can provide a 
way to measure $f(z) \sigma_8(z)$ \citep{Samushia2012}.

\begin{figure}   % GALAXY CLUSTERING -- REDSHIFT SPACE MULTIPOLES
\includegraphics[width=\columnwidth]{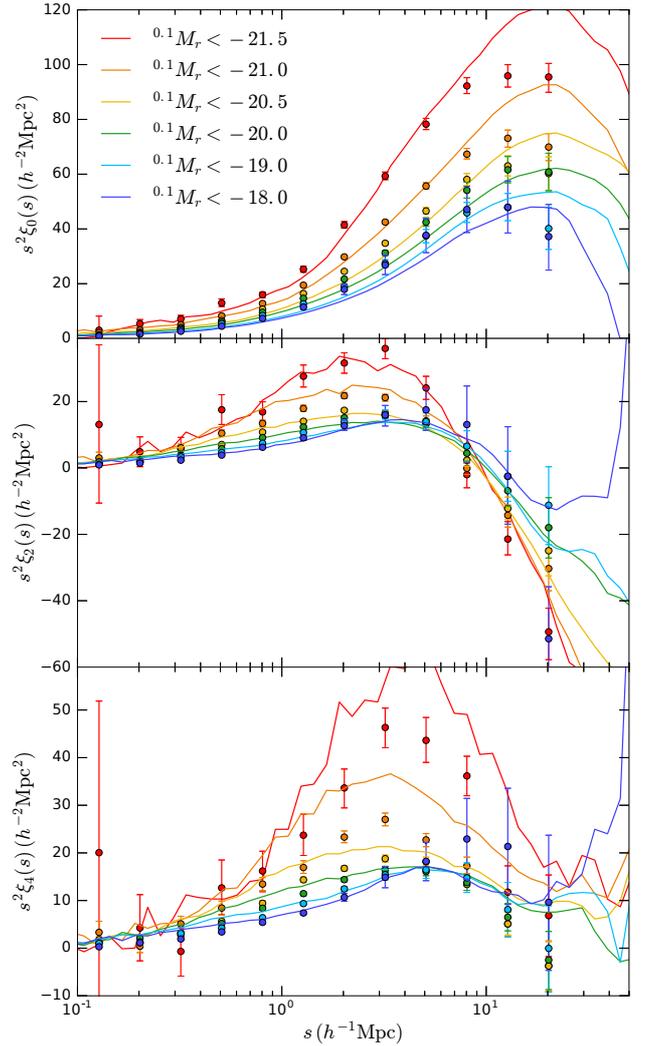}
\caption{Monopole, $\xi_0(s)$, (top panel), quadrupole, $\xi_2(s)$, (middle panel),
and hexadecapole, $\xi_4(s)$, (bottom panel), of the redshift space two-point 
correlation function for different volume limited samples (solid lines), where
the colour indicates the magnitude threshold. Points with error bars show
the measured clustering from SDSS \citep{Guo2015}.}
\label{fig:clustering_multipoles}
\end{figure}

The multipoles of the redshift-space correlation function of galaxies in the mock
catalogue are shown in Fig.~\ref{fig:clustering_multipoles} for different
volume limited magnitude threshold samples, and compared to measurements of 
clustering from SDSS \citep{Guo2015}. The monopole and quadrupole show reasonable 
agreement with the SDSS measurements, although the amplitude of the hexadecapole is a 
little high for some of the samples. 
Overall, the redshift space distortions in the mock catalogue look reasonably realistic, 
showing that the catalogue will be useful for future surveys that will take 
redshift space distortion measurements.
We have extended the predictions beyond the range of the SDSS results, where
they are easier to model and can be probed by surveys like DESI and Euclid.

%%%%% SECTION - CONCLUSIONS %%%%%

\section{Conclusions}

For upcoming galaxy surveys, such as DESI and Euclid, it is important
to have realistic mock catalogues in order to test and 
verify analysis tools, assess incompleteness and determine error covariances.
The mock catalogues can also be used to make predictions and set
expectations in advance of the first data from the survey.

We have outlined a method for creating a mock catalogue from the Millennium-XXL
(MXXL) simulation. We first created a halo lightcone catalogue from the simulation,
which we then populated with galaxies using a halo occupation distribution (HOD)
scheme. 

The halo lightcone catalogue is created from the simulation by finding the
interpolated values of the position, velocity and mass of each halo at the
redshift at which it crosses the observer's lightcone. The halo lightcone
catalogue covers the full sky, and extends to redshift $z=2.2$ with
a mass resolution of $\sim 10^{11} h^{-1}\rm M_{\odot}$. 
Extending the catalogue to high redshifts requires multiple periodic
replications of the MXXL box; these replications are only necessary
to extend to redshifts greater than $z \sim 0.5$. 

The halo catalogue is populated with galaxies using a 
Monte Carlo method, which randomly assigns galaxies with luminosities to
dark matter haloes, following a HOD. This is an extension of the 
method outlined in \citet{Skibba2006} to a 5 parameter
HOD. Galaxies in the mock catalogue are also assigned a 
$^{0.1}(g-r)$ colour, based on the Monte Carlo method of \citet{Skibba2009}. 
A galaxy is assigned a colour based on its luminosity, redshift, and whether it is 
a central or a satellite galaxy, which is randomly drawn from a parametrisation of 
the SDSS colour-magnitude diagram. 

The values of the HOD parameters we use are based on the best fitting HODs 
which reproduce the measured clustering from SDSS \citep{Zehavi2011}, but in 
Millennium cosmology. In the standard 5 parameter HOD model, the 
parameter $\sigma_{\log M}$ introduces Gaussian scatter in the luminosity
of central galaxies for haloes of a fixed mass, which can lead to
the unphysical crossing of HODs in different luminosity bins. 
We modify the HOD model so that this
scatter follows a pseudo-Gaussian spline kernel, which prevents this
crossing. The HODs are evolved with redshift such that
they reproduce a target luminosity function, which is chosen
to be the SDSS luminosity function at low redshift, and the luminosity
function from GAMA at high redshifts. For a sample of galaxies with a 
fixed number density, the shape of the HOD is kept fixed with redshift, 
and the mass HOD parameters are all multiplied by the factor $f$ which
is required to produce the correct number density. By construction,
the mock catalogue reproduces the SDSS and GAMA luminosity functions,
and the ratio of the HOD parameters $M_1/M_{\rm min}$ is constant
with redshift.

We modify the parametrisation of the colour-magnitude diagram outlined
in \citet{Skibba2009}, and add evolution, such that the distribution
of $^{0.1}(g-r)$ colours in the mock catalogue agrees with measurements
from GAMA at different redshifts.  

The galaxy catalogue is a flux limited mock galaxy catalogue,
covering the full sky with an 
{\it r}-band magnitude limit of $r<20.0$ and median redshift $z \sim 0.2$. 
The angular and projected correlation functions of galaxies in the mock 
show good agreement with measurements from SDSS
and GAMA, and the colour dependence of the clustering is reasonable. 
The BAO peak can be seen in the large scale clustering of galaxies
in the mock, and galaxies show realistic redshift space distortions,
making this mock useful for upcoming surveys which will measure
these.

Here we have presented one mock galaxy catalogue, but to enable model 
inferences and place tight constraints on cosmological 
parameters, error covariances need to be determined. This requires the 
use of many mock catalogues, of the order of 100s to 10,000s. 
This could be achieved by coupling the HOD component of the mock 
with an approximate but fast method of generating halo catalogues 
\citep[e.g.][]{Manera2013,Monaco2013,Tassev2013,White2014,Avila2015,
Chuang2015,Kitaura2015}.

\section*{Acknowledgements}

The authors would like to thank Sam Lyddon for making preliminary versions
of several of the figures, and Jon Loveday for useful discussions.
We thank the referee, Darren Croton, for several insightful comments.
This work was supported by the Science and Technology Facilities Council 
(ST/M503472/1).
AS, SMC, CMB \& PN acknowledge the support of the Science and Technology
Facilities Council (ST/L00075X/1).
ZZ was partially supported by NSF grant AST-1208891 and NASA grant NNX14AC89G.
REA acknowledges support from AYA2015-66211-C2-2.
PN acknowledges the support of the Royal Society through the award
of a University Research Fellowship and the European Research
Council, through receipt of a Starting Grant (DEGAS-259586).
IZ was supported by NSF grant AST-1612085, a CWRU Faculty Seed Grant and by 
ERC starting grant DEGAS-259586.  IZ also acknowledges the hospitality of the 
ICC at Durham University.

GAMA is a joint European-Australasian project based around a spectroscopic campaign 
using the Anglo-Australian Telescope. The GAMA input catalogue is based on data taken 
from the Sloan Digital Sky Survey and the UKIRT Infrared Deep Sky Survey. Complementary
imaging of the GAMA regions is being obtained by a number of independent survey programmes
including GALEX MIS, VST KiDS, VISTA VIKING, WISE, Herschel-ATLAS, GMRT and ASKAP 
providing UV to radio coverage. GAMA is funded by the STFC (UK), the ARC (Australia), 
the AAO, and the participating institutions. The GAMA website is 
http://www.gama-survey.org/.

This work used the DiRAC Data Centric system at Durham University, operated 
by the Institute for Computational Cosmology on behalf of the STFC DiRAC HPC 
Facility (www.dirac.ac.uk). This equipment was funded by BIS National 
E-infrastructure capital grant ST/K00042X/1, STFC capital grants ST/H008519/1 
and ST/K00087X/1, STFC DiRAC Operations grant ST/K003267/1 and Durham University. 
DiRAC is part of the National E-Infrastructure. 

%%%%%%%%%%%%%%%%%%%%%%%%%%%%%%%%%%%%%%%%%%%%%%%%%%



\bibliographystyle{mnras}
\bibliography{ref} % if your bibtex file is called example.bib

\appendix

\section{Database}
The full sky halo and galaxy lightcone catalogues are made publicly available on the
Theoretical Astrophysical Observatory database\footnote{\url{https://tao.asvo.org.au/tao/}}
\citep{Bernyk2016}. The catalogues are also available at \url{http://icc.dur.ac.uk/data/}.

\subsection{Halo catalogue}

The halo catalogue contains a total of 5.1 billion haloes out to a 
redshift of $z=2.2$, and contains the following halo properties:

\begin{itemize}
\item{$z_{\rm obs}$, the observed redshift, which takes into account the peculiar velocity of the halo}
\item{$z_{\rm cos}$, the cosmological redshift, which ignores the effect of the peculiar velocity}
\item{Right ascension, in degrees}
\item{Declination, in degrees}
\item{$M_{200 \rm m}$, the mass enclosed by a sphere in which the average density is 200 times the mean 
density of the Universe, interpolated to the redshift at which the halo crosses the lightcone, in units of 
$10^{10}h^{-1} \rm M_\odot$}
\item{$M_{200 \rm c}$, the mass enclosed by a sphere in which the average density is 200 times the critical
density of the Universe, interpolated to the redshift at which the halo crosses the lightcone, in units of 
$10^{10}h^{-1} \rm M_\odot$}
\item{$V_{\rm max}$, the maximum circular velocity, in units of $\rm kms^{-1}$}
\item $R_{V_{\rm max}}$, the radius at which $V_{\rm max}$ occurs, in $h^{-1} \rm Mpc$
\item $\sigma_{R_{\rm 200m}}$, velocity dispersion of particles within $R_{\rm 200m}$, in units of $\rm kms^{-1}$.
\item Snapshot number in the MXXL simulation
\item Halo id in the MXXL simulation

\end{itemize}

\subsection{Galaxy catalogue}

The galaxy catalogue contains 58.1 million galaxies with $r<20$, out to redshift $z=0.8$, and contains 
the following properties:
\begin{itemize}
\item{$z_{\rm obs}$, the observed redshift, which takes into account the peculiar velocity of the galaxy}
\item{$z_{\rm cos}$, the cosmological redshift, which ignores the effect of the peculiar velocity}
\item{Right ascension, in degrees}
\item{Declination, in degrees}
\item{$M_{200 \rm m}$ of the host halo, interpolated to the redshift at which the halo crosses the lightcone, 
in units of $10^{10}h^{-1} \rm M_\odot$}
\item{Apparent {\it r}-band magnitude}
\item{$^{0.1}M_r - 5\log h$, the rest frame absolute {\it r}-band magnitude, {\it k}-corrected to a reference
redshift of $z_{\rm ref}=0.1$, with no evolutionary correction}
\item{$^{0.1}(g-r)$ colour, {\it k}-corrected to a reference redshift $z_{\rm ref}=0.1$}
\item{A flag indicating whether the galaxy is a central or a satellite, and whether it is
in a resolved or unresolved halo}
\item Snapshot number in the MXXL simulation
\item Halo id in the MXXL simulation

\end{itemize}

% Don't change these lines
\bsp	% typesetting comment
\label{lastpage}
\end{document}